\documentclass{PoS}
\usepackage{epsfig}
\usepackage{graphicx}
\usepackage{amssymb}
\title{High Temperature QCD}

\ShortTitle{High Temperature QCD}

\author{{Maria Paola Lombardo} \\

        INFN \\
        E-mail: \email{lombardo@lnf.infn.it}}

\abstract{I review recent results on QCD at high temperature
on a lattice. Steady progress with staggered fermions and 
 Wilson type fermions 
allow a quantitative description of hot QCD
whose accuracy in many cases parallels that of zero temperature
studies.  Simulations with chiral quarks are coming
of age,and togheter with theoretical developments trigger interesting
developments in the analysis of the critical region. Issues
related with the universality class of the chiral transition
and the fate of the axial symmetry are discussed in the light
of new numerical and analytical results.  Transport 
coefficients and analysis of bottomonium spectra compare well with
results of heavy ion collisions at RHIC and LHC. Model field
theories, lattice simulations and high temperature systematic expansions
help building a coherent picture of the high temperature phase of QCD.
The (strongly coupled) Quark Gluon Plasma is heavily investigated,
and asserts its role as an  inspiring  theoretical laboratory.} 

\FullConference{The 30th International Symposium on Lattice Field Theory\\
		 June 24 -- 29,  2012\\
		 Cairns, Australia}

\begin{document}

\section{From here to high temperature QCD}

High temperature QCD deals with phenomena related with  the deconfinement
transition towards Quark Gluon Plasma, and the Quark Gluon Plasma phase 
itself\cite{Jacak:2012dx}. 

When heating up matter  from room temperature towards 
temperatures which are best measured in MeV rather than in 
ordinary degrees  a gas of the lightest particles --
a gas of pions -- is first produced. 
As temperature increases, a more general gas of resonances,
the Hadron Resonance Gas, 
is though to be a better approximation of the system. 
Beyond temperatures corresponding to the nucleosynthesis of the lightest nuclei we enter what we can consider a purely hadronic world. At a 
temperature of about 200 MeV -- 
$ \simeq 2\dot 10^{12} $ K -- hadrons cannot exist anymore, and a transition to  
the Quark Gluon Plasma takes place. The plasma
survives  till the Electroweak transition, at a temperature of 
O$(100)$ GeV. All in all, the field of high temperature QCD
spans about five orders of magnitude in energy, 
between the nucleosynthesis of light nuclei up to the
Electroweak transition.   As we -- lattice theorists -  know,
many of the high temperature most significant manifestations 
are outside
the reach of any analytic approach, making lattice studies mandatory. 

This paper describes the developments of this field after last year's 
Lattice review
\cite{Levkova:2012jd}. 
The material  fits roughly in two main avenues : one more
phenomenological line, which uses QCD, or approximation thereof, to describe
mostly the physics of high temperature around the critical region, 
up to temperatures of about 600 MeV. High statistics, physical 
values of the quark masses,
continuum limit are key ingredients here. One second line
uses models or approximate calculations to learn about general 
aspects of the same  phenomena. This second line reaches up to far higher 
temperatures, witnessing the deep theoretical interest of the yet 
largely unexplored physics of the Quark Gluon Plasma.

Before closing this short introduction, one word about the experimental 
situation. During the first years of operation at RHIC there was a strong focus and  motivation to explore the physics of small baryon densities -- which, in 
the Gran Canonical approach we use, corresponds to the physics of 
$\mu/T  < 1.$, $\mu$ being the quark chemical potential. The current heavy ion collisions at the  LHC  create instead near zero density systems, $\mu/T << 1 $ 
while the search 
of the elusive endpoint of QCD and the study of other aspects of colder, 
and denser matter $\mu/T > 1$ 
will be investigated at RUN II at RHIC, and future 
experiments at FAIR and NICA. 
This review will concentrate then on the first aspect -- the
physics of zero baryon density. The current effort at non--zero baryon density,
towards the understanding of the difficult, technical problems still hampering
the simulations at large chemical potential,  is discussed by  
Gert Aarts \cite{Aarts}.

\section{Thermodynamics : from hadronic Phase to QGP}
The transition from the hadronic phase to the Quark Gluon Plasma
is a crossover for physical values of the quark masses\cite{Levkova:2012jd}. 
As such, the transition temperature is not uniquely defined and in 
past years a considerable effort has been devoted to accurate measurements
of the pseudocritical temperature identified by different observables. 
With staggered fermions there is now agreement among different estimates,
when the same observables are being used.  A key ingredient in 
resolving this issue has been a careful scale setting procedure. 
\begin{figure}
\hskip 1.5truecm
\begin{minipage}{\textwidth}
\vskip -9 truecm
\hskip -2.0 truecm
 \includegraphics[width=0.8\textwidth]{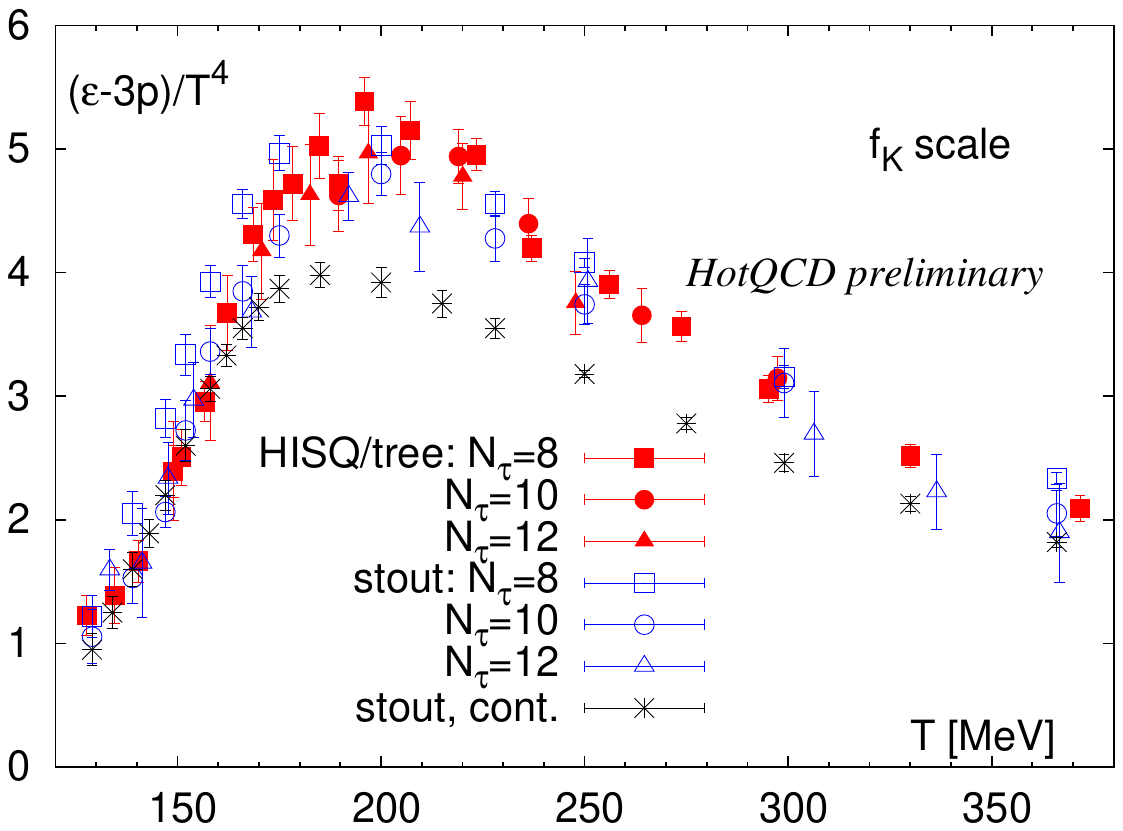}\hskip -5.5 truecm\includegraphics[width=0.8\textwidth]{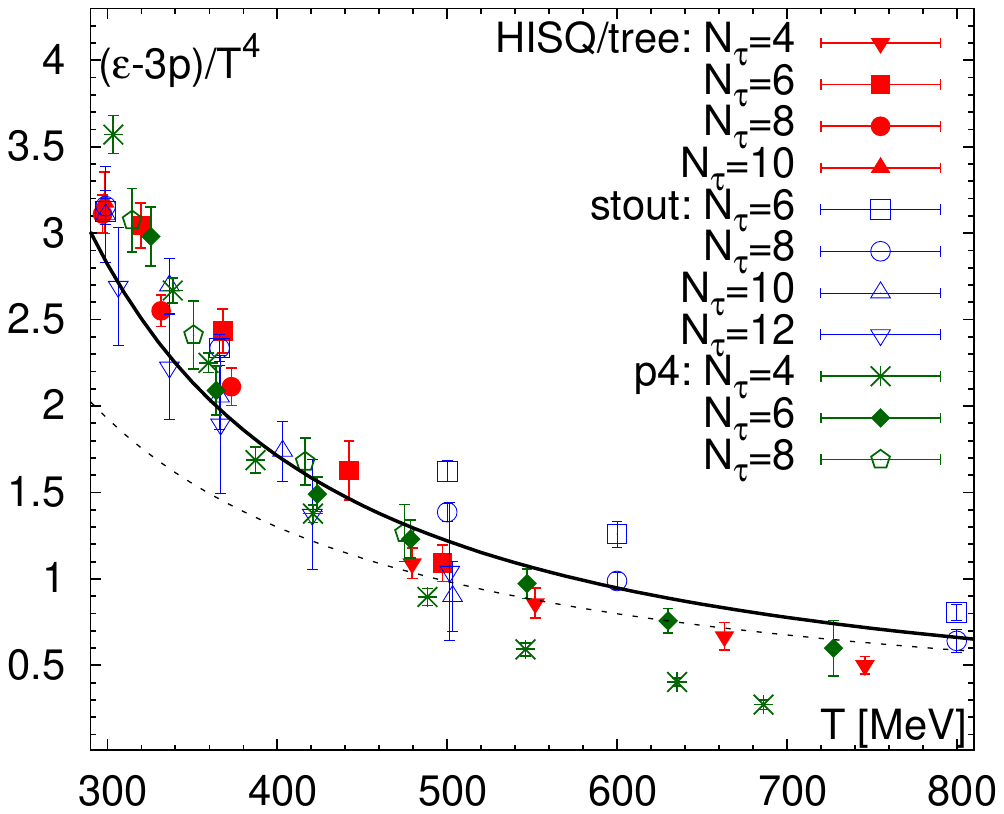}
\end{minipage}
\caption{Equation of state for
$N_f=2+1$ flavours, staggered quarks. The 
HotQCD collaboration use HISQ/tree action, the Wuppertal--Budapest
a stout action. The pion mass in either cases is close to its
physical value. 
Note the different temperature ranges in the two diagrams 
(Ref. \cite{Petreczky}).}
\label{fig:combopeter}
\end{figure}
Still not completely converged before Lattice 2012
was  the situation of the thermodynamic equation of state,
namely the relation between pressure, energy and temperature. 
The phenomenological relevance is great, since the equation of state
is an
input to hydrodynamic calculations \cite{CasalderreySolana:2011us}. 
The quantity
which is more easily accessible in a lattice calculation is
the trace anomaly $\epsilon - 3P$ (often called interaction
measure) which can be computed by differentiating the 
partition function \cite{DeTar:2009ef}. 
It is customary to compare its behaviour
to that of the Stefann-Boltzmann limit, corresponding to free fields:
convergence to this limit is achieved only at very large temperatures,
$T > 10^{18}$ GeV. This late convergence is taken as one of the evidences of
the interacting nature of the Quark Gluon Plasma close to the
critical temperature. 

\subsection{Staggered fermions with physical values of u,d,s masses}
Only two collaborations have results with physical values
of the up,down, strange mass, in either cases with staggered fermions.   
In literature and talks prior to Lattice 2012 
there are reports of a
discrepancy between HotQCD results on 
the interaction measure $\epsilon-3P$ and the
Wuppertal--Budapest stout continuum estimate.
 
\begin{figure}
\centering 
\hspace{-2mm}
\includegraphics[width=9.0cm,height=5.5cm]{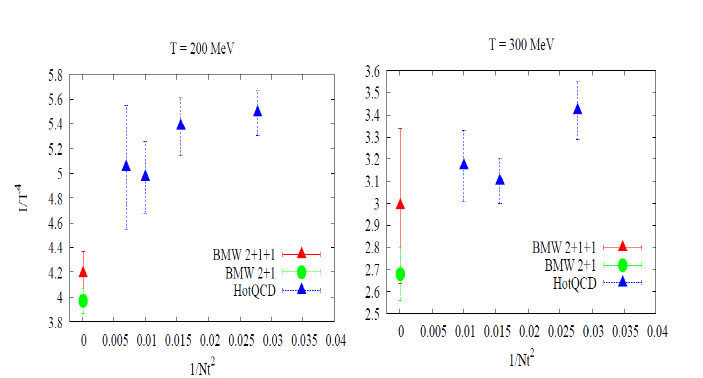}
\includegraphics[width=5.0cm,height=5.0cm]{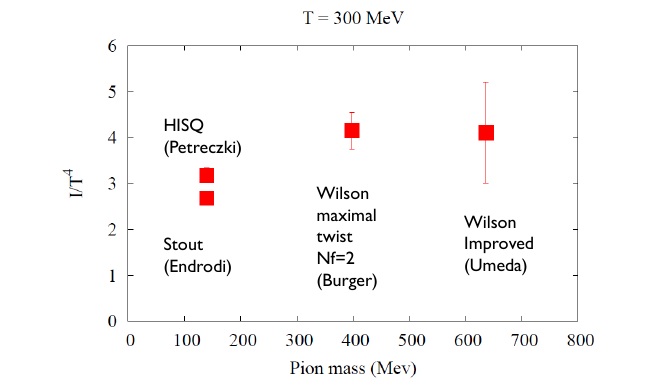}
\caption{The interaction measure for two selected temperatures as a function
of the $1/N_t^2$ : results at fixed $N_t$ compare favourably, some residual
issues with the extrapolation procedure are being settled. On the right,
the overall status of the continuum results for $T=300 $ MeV (Numerical results communicated by the Authors).}
\label{fig:compare}
\end{figure}

In Fig. \ref{fig:combopeter}, from Ref.\cite{Petreczky}, 
we can see the plot of the trace anomaly $(\epsilon - 3P)/ T^4$  
as a function of 
the physical temperature, as obtained from the two major 
staggered collaborations
using stout\cite{Borsanyi:2012cr} and HISQ staggered 
quarks\cite{Petreczky}   and physical quark up,down and strange masses.
Once the scale is properly set by using  $f_K$,
for a given value of $N_t$ the results  nicely
agree with each other. The extrapolated data -- according to the
Wuppertal-Budapest collaboration -- however lie significantly below
the results for any finite $N_t$, suggesting residual 
discretization effects. 
As thermodynamics is now entering
a precision era, perhaps it is useful to fix some significant values of the
temperature and compare in detail results -- similarly to what we would
do at $T=0$. I have chosen $T=200$ MeV -- in correspondence to the peak
of the interaction measure -- and  $T = 300$  MeV, see 
Fig. \ref{fig:compare},left. 
Since for a fixed $N_t$ the agreement is
very good, I only show the finite $N_t$ HISQ results, together with 
the stout extrapolated  results.   The relevance of the continuum 
extrapolation appears clearly in the plots -- the extrapolation
of the HISQ result depends crucially on which $N_t$ one would include.
In this sense, the resolution of the residual small discrepancy
is narrowed down to the issue of $N_t$ extrapolation,  and either 
collaborations  are steadily improving their results. 
The same Fig.\ref{fig:compare}, right,  shows the dependence 
of bulk thermodynamics on the matter content, to be discussed below. 

\subsection{Wilson fermions thermodynamics}
New results for the thermodynamics with Wilson fermions
have been presented by the WHOT--QCD collaboration\cite{Umeda} and  
by the tmft  collaboration \cite{Burger}. In either cases
the pion mass is larger than $400$ MeV.

The WHOT--QCD collaboration  presented results for $2+1$ flavors,  
a physical value of the strange quark mass, and a largish pion mass.
Some results are shown in Fig. 3.
 The Wilson action was non--perturbatively improved, and the gluonic
part used  Iwasaki improved glue.  
Results for two flavors of twisted mass Wilson fermions at maximal twist 
are available as well, with  pion masses of about 400 MeV, 
and larger, see Fig.4. 
 The results used the tree--level
correction advocated by the Wuppertal Budapest collaboration, and 
the extrapolation to the continuum
limit appears to be reasonably under control. 
\begin{figure}
\begin{minipage}{7.7cm} 
\vspace{0.6cm}
\includegraphics[width=6truecm,height=4.5cm]{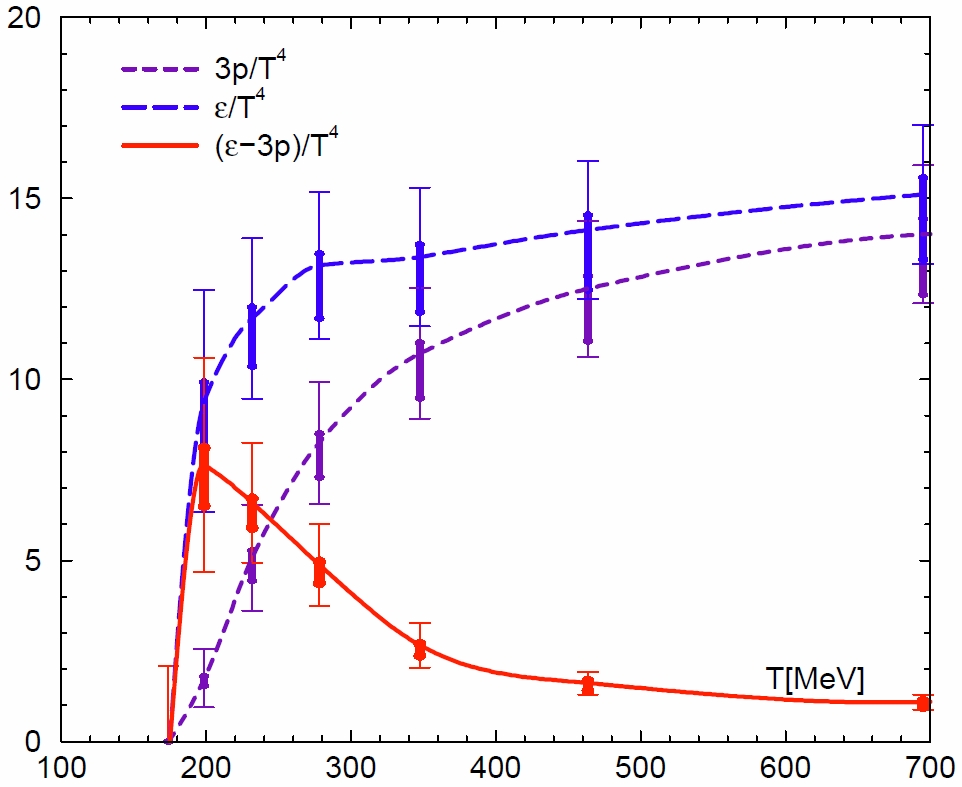}
\label{fig:umeda}
\vspace{0.6cm}
\caption{Equation of state for $N_f=2+1$ 
improved Wilson fermions.
$m_\pi/m_\rho = 0.63$ (Ref. \cite{Umeda}). 
} 
\end{minipage}
\hspace {0.4cm}
\begin{minipage}{7.7cm}
\includegraphics[width=6truecm,height=5.7cm]{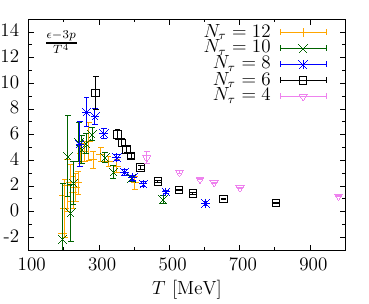}
\caption{Equation of state for $N_f=2$ Wilson  fermions at maximal twist, 
$m_\pi \simeq 400$ MeV (Ref. \cite{Burger}).}
\label{fig:florian_eos}
\end{minipage}
\end{figure}
\subsection{Effects of the charm on the EoS}
\begin{figure}
\centering
\includegraphics[scale=0.25]{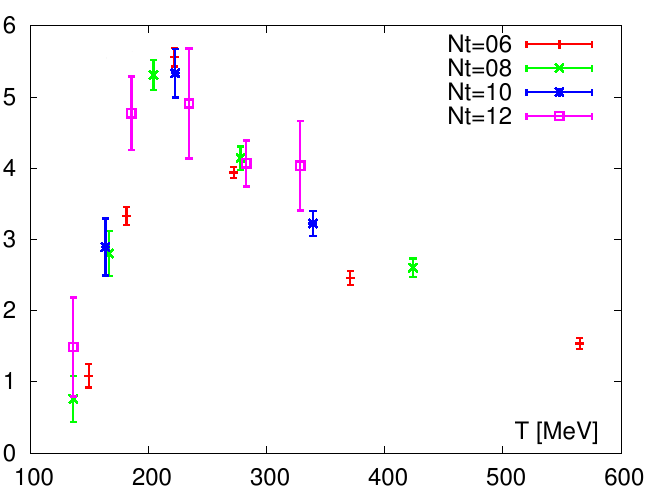}
\hspace{0.3cm}
\includegraphics[scale=0.25]{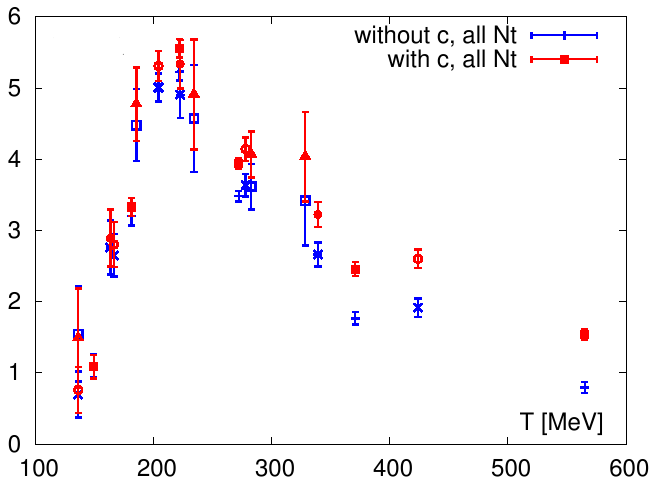}
\caption{Charm effects on the Equation of State: 
MILC results for the trace anomaly for 
$N_f=2+1+1$ using HISQ quarks  
along lines of constant physics(left) and comparison of results
with and without dynamical charm(right). This study was carried out
with physical charm and strange masses. The comparison of
the left and right plots allows the identification of the effect of
the charm mass on the EoS (Ref. \cite{Heller}).}
\label{fig:milc}
\end{figure}
One step ahead in the matter content was taken by the MILC  
collaboration and by the Wuppertal-Budapest Collaboration who presented 
results for $N_f=2+1+1$. 
Contrary to  naive expectations, and
in agreement with the predictions of effective theories\cite{Laine:2006cp}, 
there is an early onset
of the charm contribution at about $T = 350$ MeV.
The MILC collaboration\cite{Heller} determined
the QCD EoS with $2+1+1$ flavors using the HISQ action.
Their 
study was then carried out  on lines of constant physics with $m_l/m_s = 1/5$,
with physical strange and charm quark masses. Above $T_c$ the precise value
of the light quark mass should not be important.  The 
preliminary data have been obtained with $N_t$=6, 8, 10 and 12, 
without any  attempt at a continuum extrapolation yet.
Fig.\ref{fig:milc}, left,  shows the  MILC preliminary results 
for the trace anomaly $\epsilon - 3P$ , and in Fig.\ref{fig:milc}, right, 
the same is shown  but with either without 
the valence charm quark contribution.

The Wuppertal-Budapest collaboration\cite{Krieg} presented 
further results on charm effects on the EoS. Their results
 were also successfully compared with High Temperature  perturbation theory, 
down to $T = 350$  MeV, and with a Hadron Resonance Gas
model, which offers a good description of 
the system up to temperature of about $150$ MeV. 
The outcome of this comparison is presented in Fig.\ref{fig:krieg}.
As discussed at length in previous studies, any simple analysis is expected
to break down in the vicinity of the transition, where 
the critical or pseudocritical behaviour cannot be captured by an analytic 
equation of state. Fig. \ref{fig:krieg} thus clearly identifies the
interesting region where a lattice analysis is mandatory. 
An overview of the mass sensitivity for $T=300$  MeV
can be seen in  Fig.\ref{fig:compare}, right, where all available
continuum results 
are shown as function of the pseudoscalar mass. 
\begin{floatingfigure}
\includegraphics[width=0.5 \textwidth]{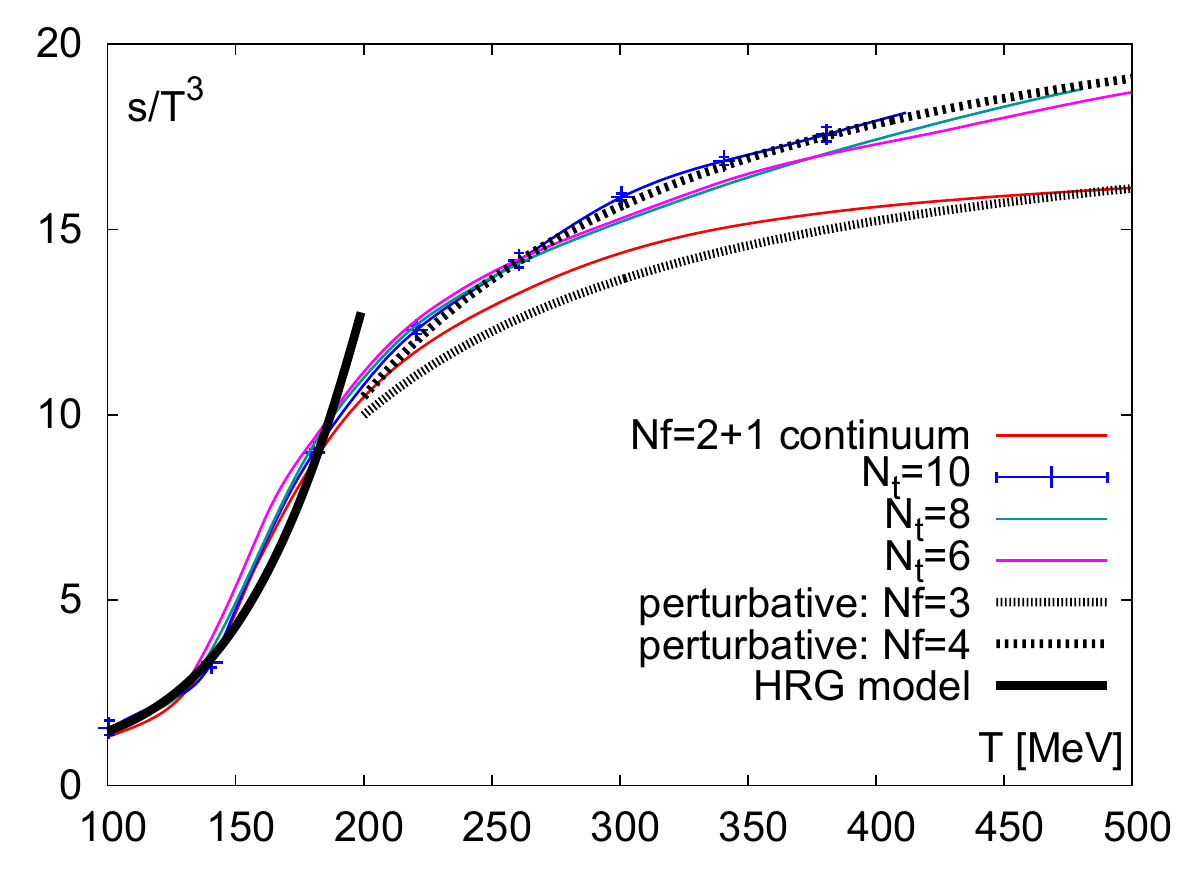}
\caption{The entropy for the Nf=2+1+1 flavors from the Wuppertal-Budapest group:the low temperature results compare well with the HRG model, while high temperature
results demonstrate the sensitivity to the charm in the perturbatively predicted region (Ref. \cite{Krieg}).}
\label{fig:krieg}
\end{floatingfigure}

\section{The transition region -- symmetry aspects}
In the limit of massless quarks 
the QCD Lagrangian enjoys a global chiral symmetry
$SU(N_f)_L \otimes SU(N_f)_R \otimes U_A(1) \otimes U_V(1)$. 
The vector $U_V(1)$ transformation 
corresponds to baryon number and the axial $U_A(1)$ which would promote the
symmetry to $U(N_f) \otimes U(N_f)$ is a symmetry of the classical theory but
not of the quantum theory. The symmetry breaks down to 
$SU(N_f) \otimes U_V(1)$ at low temperatures. Since up and down quarks, and to
some extent the strange quarks, are not too heavy, relics of the
critical behaviour can still be observable even in the crossover
region for physical values of the quark masses - the connection  being
made within the framework of the magnetic equation of
state. 

It is speculated that the breaking of the $U_A(1)$ 
symmetry is stronger at zero temperature, or, equivalently, 
that $U_A(1)$ symmetry
is effectively restored at high temperature. 
The fate of the $U_A(1)$
effective restoration is closely related with the issue of the
order of the chiral transition with two massless flavors,
leaving open the possibility of either a first or second order transition
belonging to different universality classes, 
depending on the intensity of the residual $U_A(1)$ breaking at 
$T_c$ \cite{Pisarski:1983ms,Butti:2003nu,Vicari:2007ma,Vicari:2008:jv}.  

 In view of the relevance of the symmetries, it is 
obviously extremely desirable 
to have fermionic formulations which respect chiral symmetry. 
Remember now that staggered and Wilson fermions break the $U_A(1)$ 
symmetry, and that chiral symmetry is also broken in the Wilson
formulation, while a $U(1)$ symmetry (unrelated with the
$U_A(1)$ symmetry)  is preserved by staggered
fermions, which however do not allow a local definition of $U_A(1)$
symmetry. The computationally expensive overlap quarks have an exact 
chiral symmetry. Domain wall fermions approximate it with a  precision 
which is  in principle  arbitrary, the residual chiral symmetry breaking
being characterised by an additive renormalisation to the quark mass
$m_{res}$ which vanishes in the limit of infinite 5-th dimension:  
$m_{res} (L_s)  = c_1 \frac{e^{-\lambda L_s}}{L_s} + c_2 \frac {1}{L_s}$. 

New results with Wilson fermions with $2+1$ and $2$ flavours, as well as
results with chiral fermions were presented at this conference, with
emphasis on  two interrelated open issues: the fate of the $U_A(1)$ symmetry
at the phase transition, and the universality class for $N_f=2$. 

\subsection{$U_A(1)$}
The physics of the anomaly is particularly intriguing, given its role
in the universality class of the 
transition
\cite{Pisarski:1983ms,Butti:2003nu,Vicari:2007ma,Vicari:2008:jv}. 
Moreover, should the anomaly
not be restored at the transition, 
one would have the opportunity to disentangle
its effects from those of chiral symmetry.

Patterns of symmetries are reflected 
in the interplay of 
scalars susceptibilities, according to the well known MILC plot
\cite{Shuryak:1993ee,Bernard:1996iz}. 
The order parameter for $SU(2) \otimes SU(2)$ is
related to the longitudinal susceptibility 
$<\bar \psi \psi> / m = \chi_\pi$ , and a candidate
 order parameter  for $U_A(1)$ is $\chi_\pi - \chi_\sigma$. 
It is however true in general
that if an observable $\cal O$ is not invariant under a given symmetry $\cal S$,
$<{\cal O}> = 0$ is a necessary, but not sufficient
condition for $\cal S$  to be exact. This
means that  the
candidate order parameter for the $U_A(1)$ 
symmetry $\chi_\pi - \chi_\sigma$  might well be zero even if
$U_A(1)$ is still broken. %

Chiral observables are related to the low-lying eigenmodes of Dirac operator:
\begin{eqnarray}
<\bar \psi \psi>& = & \int d \lambda \rho(\lambda) 
\frac {2 m}{m^2 +\lambda^2}\\
\chi_\pi - \chi_\delta & = & \int d \lambda \rho(\lambda) 
\frac {4 m^2}{(m^2 +\lambda^2)^2}
\end{eqnarray}
 so the fate of the $U_A(1)$
symmetry can be read off the shape of the eigenvalue density $\rho(\lambda)$.
The various possibilities are 
extensively discussed in a recent paper by the
HotQCD Collaboration \cite{:2012jaa}.

A new theoretical analysis has been presented by 
Aoki
and collaborators\cite{Aoki,Aoki:2012yj}, who find
\begin{equation}
 \lim_{m \to 0} < \rho^A(\lambda)>_m = \lim_{m \to 0} <\rho_3^A>_m \frac{\lambda^3}{3!} + O(\lambda^4)
\end{equation}
i.e. no constant contribution to the spectrum,  
which implies at high temperature 
\begin{equation}
\lim_{m \to 0} (\chi_\pi - \chi_\sigma) = 0.
\end{equation}
when $<\bar \psi \psi> = 0$.  In other words, 
for any $N_f$,  two point correlators -- hence susceptibilties -- 
would be blind
to the breaking of $U_A(1)$ when chiral symmetry is restored. 
Note that
older results (refs. \cite{Lee:1996zy,Cohen:1996sb}) 
reached a similar conclusion on the non-observability
of $U_A(1)$ non restoration in two point functions, 
but only for $N_f \ge 3$. The new analysis \cite{Aoki} 
requires a careful consideration
of the order of the limits and suggests that early work
was inaccurate. 

\begin{figure}
\includegraphics[width=7.5truecm,height=4.5cm]{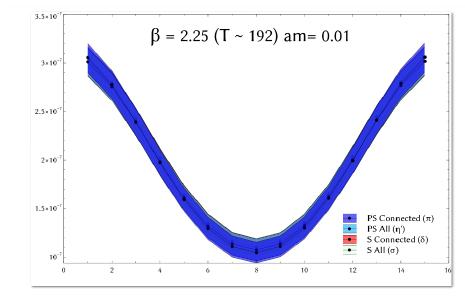}
\includegraphics[width=7.5truecm,height=4.3cm]{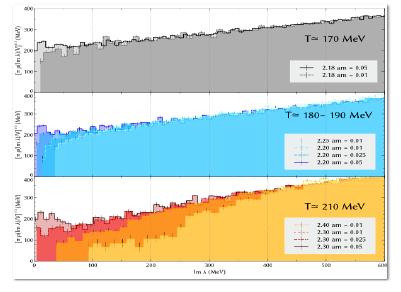}
\caption{Direct observation of the degeneracy in the scalar sector, right
above $T_c$, left. The eigenvalue distribution for several temperatures,
hinting at the restoration of the $U_A(1)$ symmetry, right. Overlap fermions, 
two flavors (Ref.\cite{Cossu}). }
\label{fig:guido}
\end{figure}

Results presented by JLQCD\cite{Cossu}, 
with two flavors of overlap fermions 
and physical values of the pion mass,
apparently realise the  scenario proposed by Aoki\cite{Aoki} 
indicating the degeneracy of 
all the scalar correlators in the chiral limit above $T_c$, which is
of course a stronger statement than the degeneracy of the susceptibilities,
see Fig. \ref{fig:guido}, left. 
Moreover, a gap in the eigenvalues' distribution opens up right above $T_c$,
as shown in Fig. \ref{fig:guido}, right. 
The conclusion still needs 
to be corroborated by a better control of systematics.

\begin{figure}
\begin{minipage}{0.5 \textwidth}
{
\includegraphics[width=8truecm,height=3.7cm]{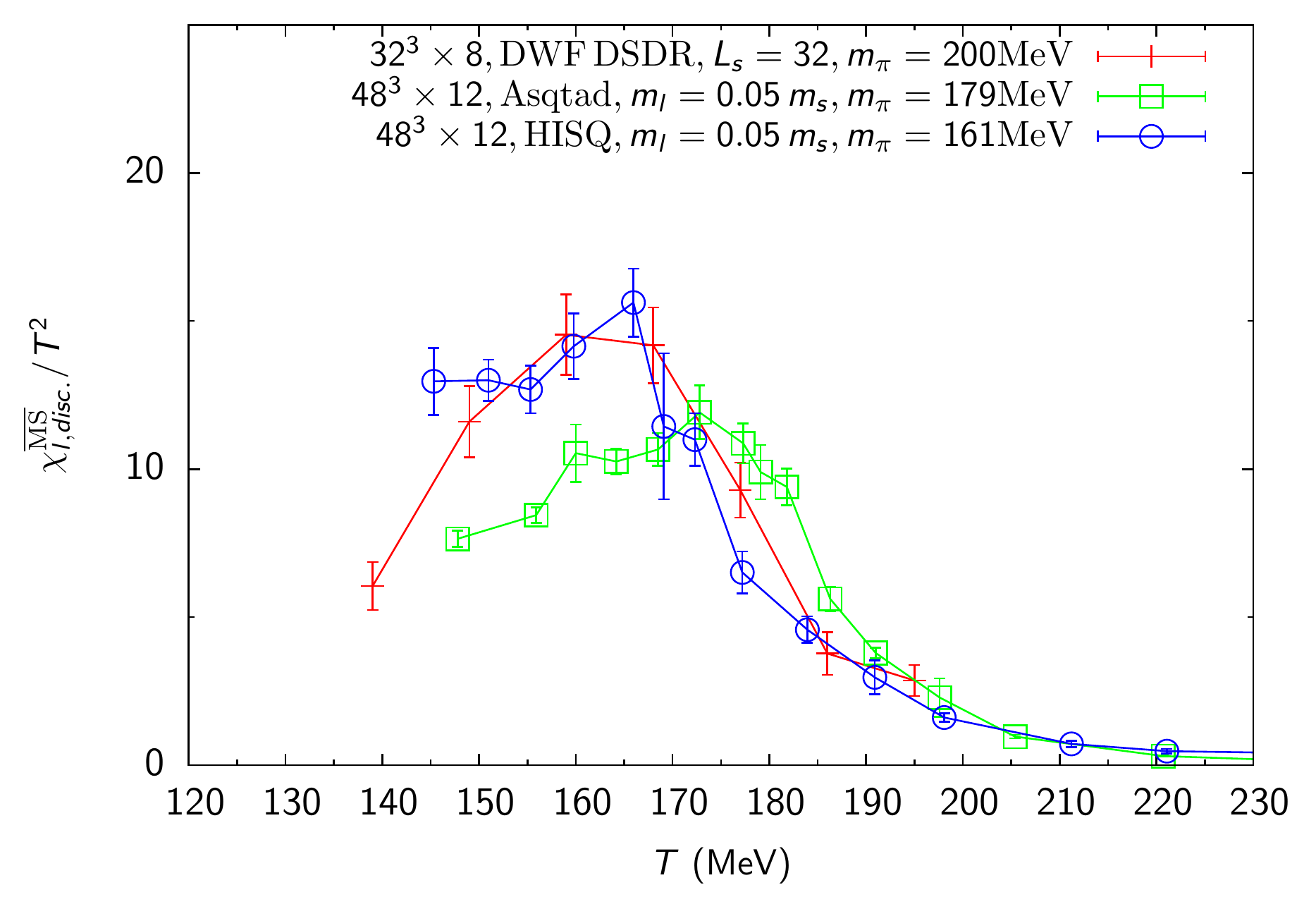}\\
\includegraphics[width=8truecm,height=2.3cm]{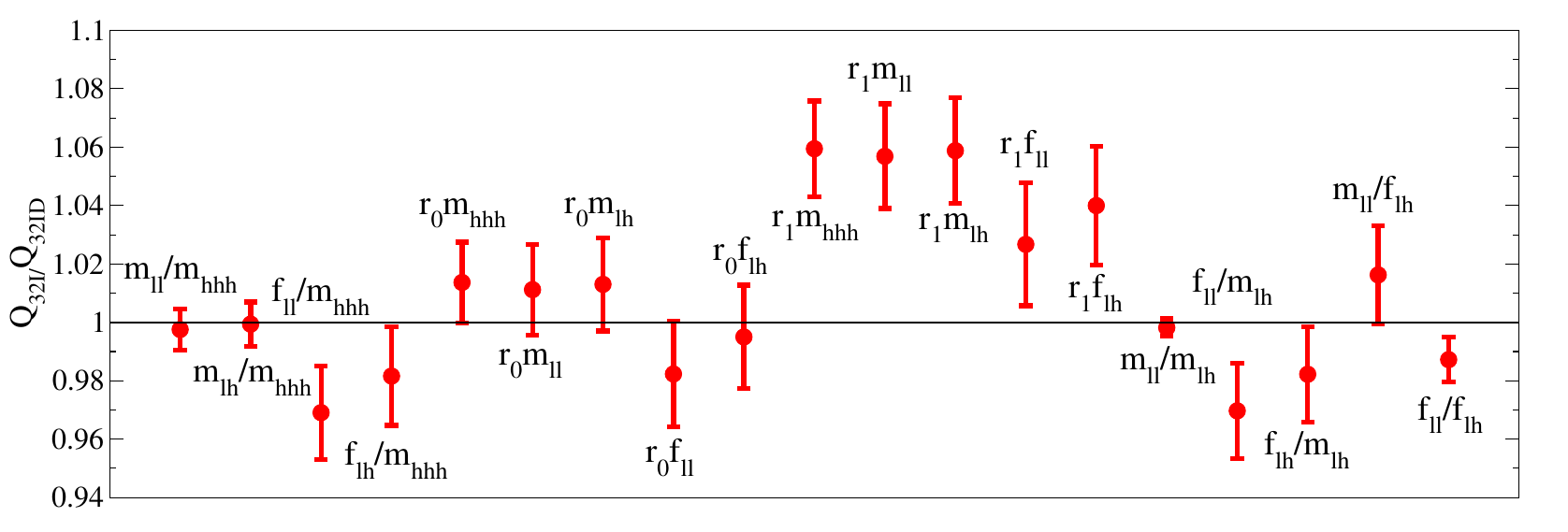}}
\end{minipage}
\hspace{0.3cm}
\begin{minipage}{0.5 \textwidth}
\vspace{0.5cm}\includegraphics[width=7.5truecm,height=7.2cm]{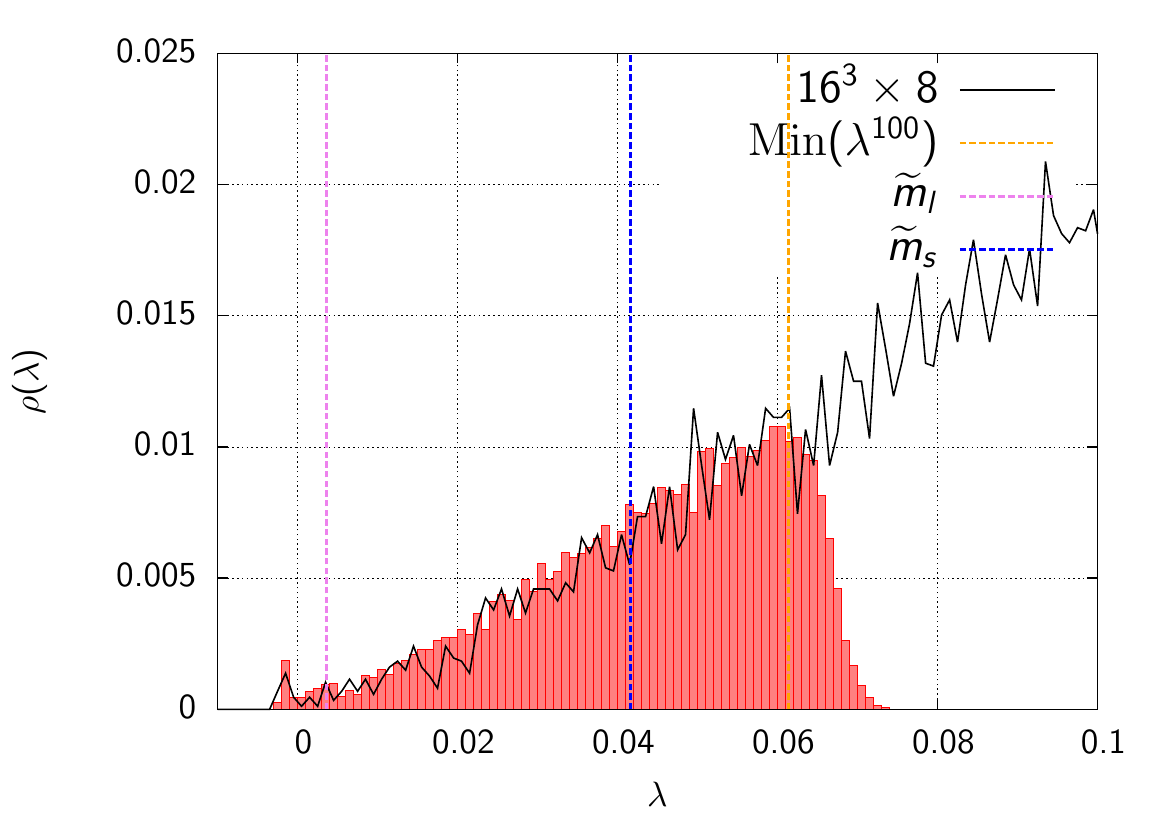}
\end{minipage}
\caption{ Domain wall fermions via-a-vis 
staggered results(top left).
The scale independence for domain wall fermions (bottom left).
The eigenvalue spectrum (right)   favours
a non restored $U_A(1)$ symmetry beyond the QCD transition, however
the largish masses and the residual breaking leave room for
a different conclusion (Ref. \cite{Lin}).}
\label{fig:DiracDWF}
\end{figure}

The same issue was investigated with  domain wall fermions.  
New results with $2+1$ flavours 
were presented at this meeting\cite{Lin}
on the $32^3 \times 8$ lattices by RBC/LLNL, and ensembles with physical
quark masses on the $32^3 \times 8$ and $64^3 \times 8$ 
are being planned from HotQCD. 
The behaviour of chiral susceptibilities which can be seen in 
Fig.\ref{fig:DiracDWF}, left top, is on the overall convincing,
with  small deviations from the existing staggered results. 
Fig.\ref{fig:DiracDWF},left bottom, shows the ratio of dimensionless
ratios at $\beta=1.75$,  demonstrating   
violation of the order of only 5 \%  from an ideal scaling. 
The spectral densities are shown in 
Fig.\ref{fig:DiracDWF}, right, for a temperature 
above, but  close to $T_c$.
The distribution shows a clear peak near zero 
("near zero modes") and a linear behaviour extending
down to zero.  This may correspond in the continuum 
limit to $\rho(\lambda) = c_1 \delta(\lambda)+c_2\lambda.$
at a variance with Aoki's scenario, and 
implying a  residual $U_A(1)$ breaking with 
$\chi_\pi-\chi_\delta \ne 0 $ in the chiral limit.

All these analysis 
require an extrapolation to the chiral limit, hence 
studies with different masses -- which are in progress - 
are needed to  draw definite conclusions.
However a detailed study with HISQ fermions\cite{Ohno}, $2+1$ flavors, 
pion masses
ranging from $115$ MeV till $230$ MeV, leaves the conclusion open, 
with a careful discussion of the systematic effects affecting a proper
analysis of the eigenvalue shape. The results for the distributions are
presented in Fig. 9. 
Even at the highest temperature
there is no evidence of a gap around zero.  
More than any comment the shape of the
eigenvalues clearly demonstrate the uncertainties affecting the conclusions.

\subsection{Further studies around the critical temperature}
Further studies with chiral fermions were presented by the TWQCD
Collaboration , using two flavours of Domain Wall fermions \cite{Hsieh}
 and by the Wuppertal-Budapest 
collaboration\cite{Krieg,Borsanyi:2012xf}, who use
two flavours of overlap fermions, with topology fixing \'a  la JLQCD. 
These studies should be regarded as still preparatory since 
the pion masses exceed $300$ MeV, 
and the analysis concentrates on  basic observables. 
The Wuppertal-Budapest collaboration
presented --see Fig.\ref{fig:overwup} --  a comparison with  continuum
staggered results which  is at first sight 
satisfactory\cite{Krieg,Borsanyi:2012xf}. 
\begin{figure}
\begin{minipage}{\textwidth}
\vskip -3cm
\hskip -1cm
\includegraphics[width=11truecm]{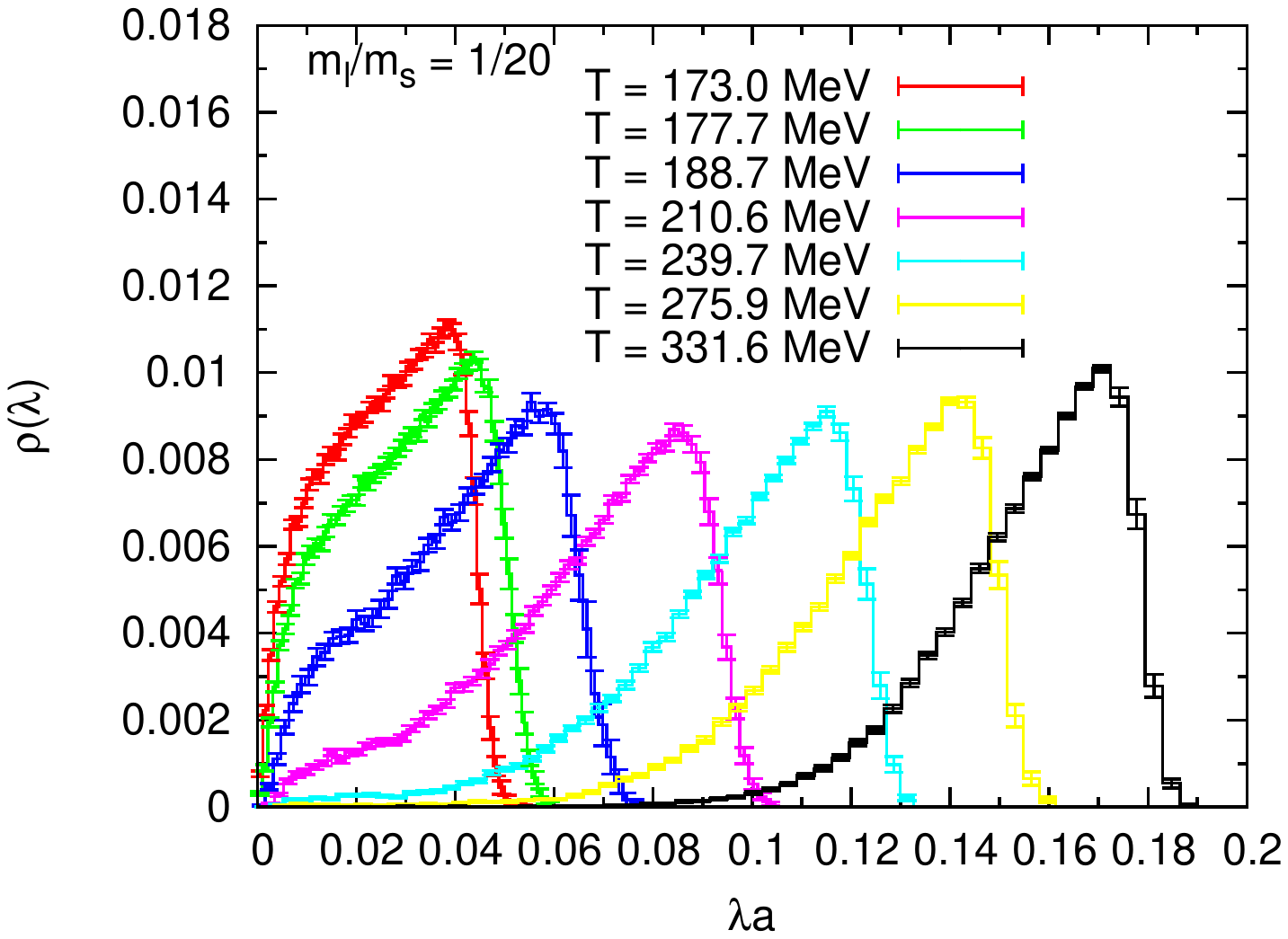}
\hskip -4cm 
\includegraphics[width=11truecm]{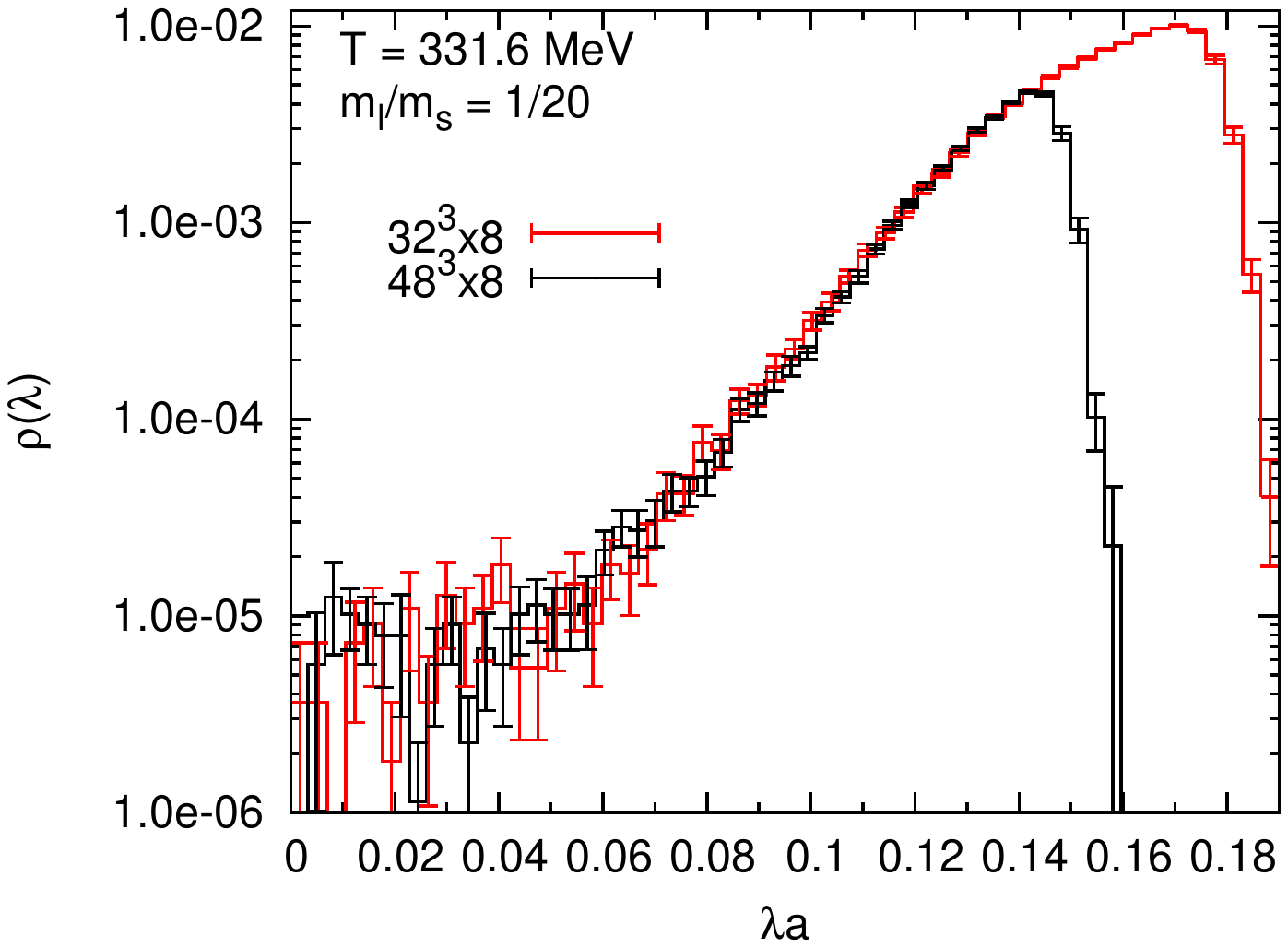}
\label{fig:Ohno}
\vskip -1cm
\caption{Eigenvalue distribution from HISQ fermions: the shape determines
the fate of $U_A(1)$ symmetry which remains open.The left diagram 
shows the temperature dependence of the Dirac spectrum for
temperatures in the range $173 - 332$ MeV $> T_c$. 
The right diagram shows the volume dependence
of spectral density at the  highest
temperature. The volume dependence is quite small and there
remains a tail even at this temperature (Ref. \cite{Ohno}).}
\end{minipage}
\end{figure}
The transition region 
was also studied with Wilson fermions, $2+1$ flavors, including continuum
extrapolation\cite{Nogradi,Borsanyi:2012uq}. 
The pion masses range from $125$ to $275$ MeV, and 
studies with smaller masses are in progress. 
Again the focus is on the agreement
with staggered quarks, see Fig. 11, 
which so far looks very nice,  within the largish errors.
\begin{figure}
\centering
\includegraphics[width=7.7truecm,height=4cm]{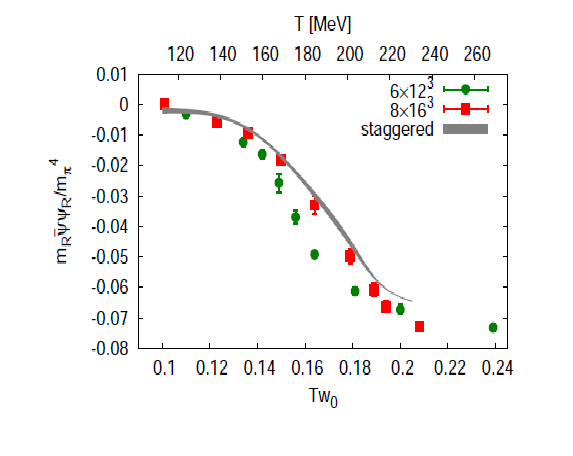}
\hspace{-0.9cm}\includegraphics[width=7.7truecm,height=4cm]{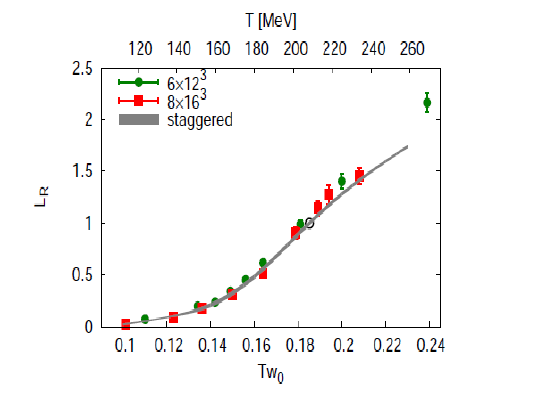}
\caption{The crossover region with overlap fermions, $N_f=2$,
 $m_\pi = $ 350 MeV (Ref. \cite{Borsanyi:2012xf}).}
\label{fig:overwup}
\end{figure}
\begin{figure}
\includegraphics[width = 4.5 truecm]{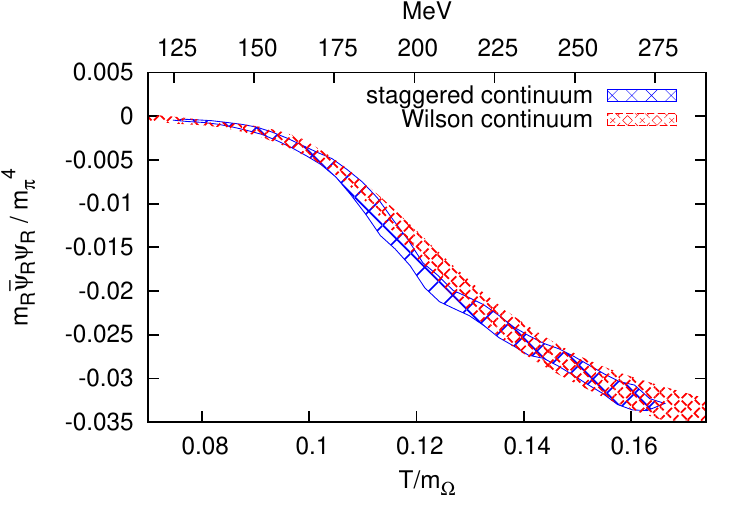} 
\includegraphics[width = 4.5 truecm]{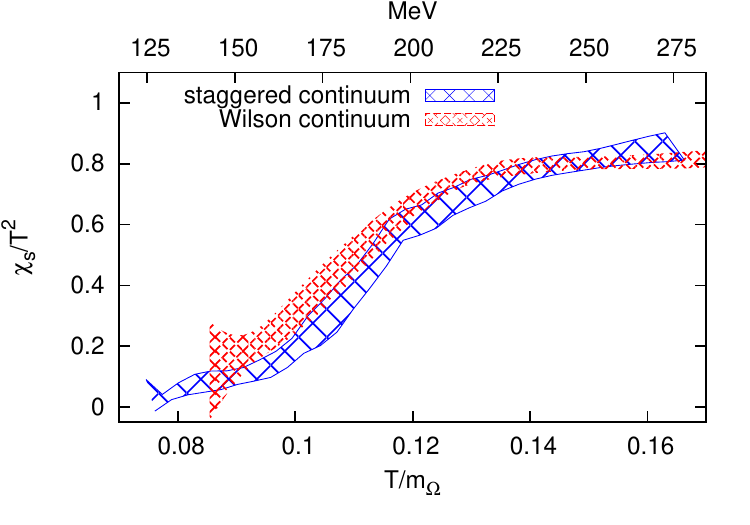}  
\includegraphics[width = 4.5 truecm]{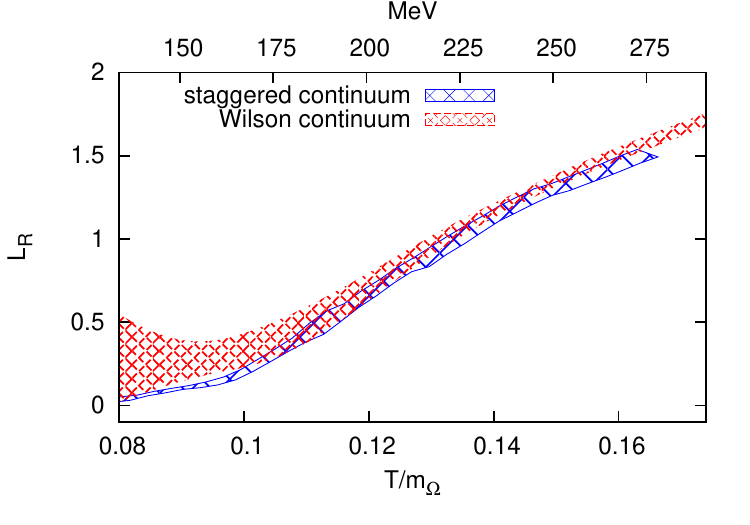}
\label{fig:Nogradi}
\caption{The renormalised chiral condensate, renormalised Polyakov loop and strange quark susceptibility from the Wuppertal-Budapest simulations with 
2+1 flavors of Wilson fermions (Ref. \cite{Nogradi}).}
\end{figure}
New  results for two flavors of non-perturbatively 
$O(a)$ improved  Wilson fermions 
have been presented\cite{Brandt}, 
 working along lines of constant physics
with $m_\pi \simeq  290$  MeV.  The picture, Fig. 12,
 shows the axial vector correlator
normalised to the free vector correlator. 
The right-side picture shows the collection of screening masses
themselves - the degeneracy in the vector sector which anticipates the
one in the scalar sector might suggest a persistence of $U_A(1)$ breaking,
although it could also be that the near degeneracy in the vector sector comes
from the largish masses' values.
\begin{figure}
\centering
\includegraphics[width=6.4truecm,height=4.7cm]{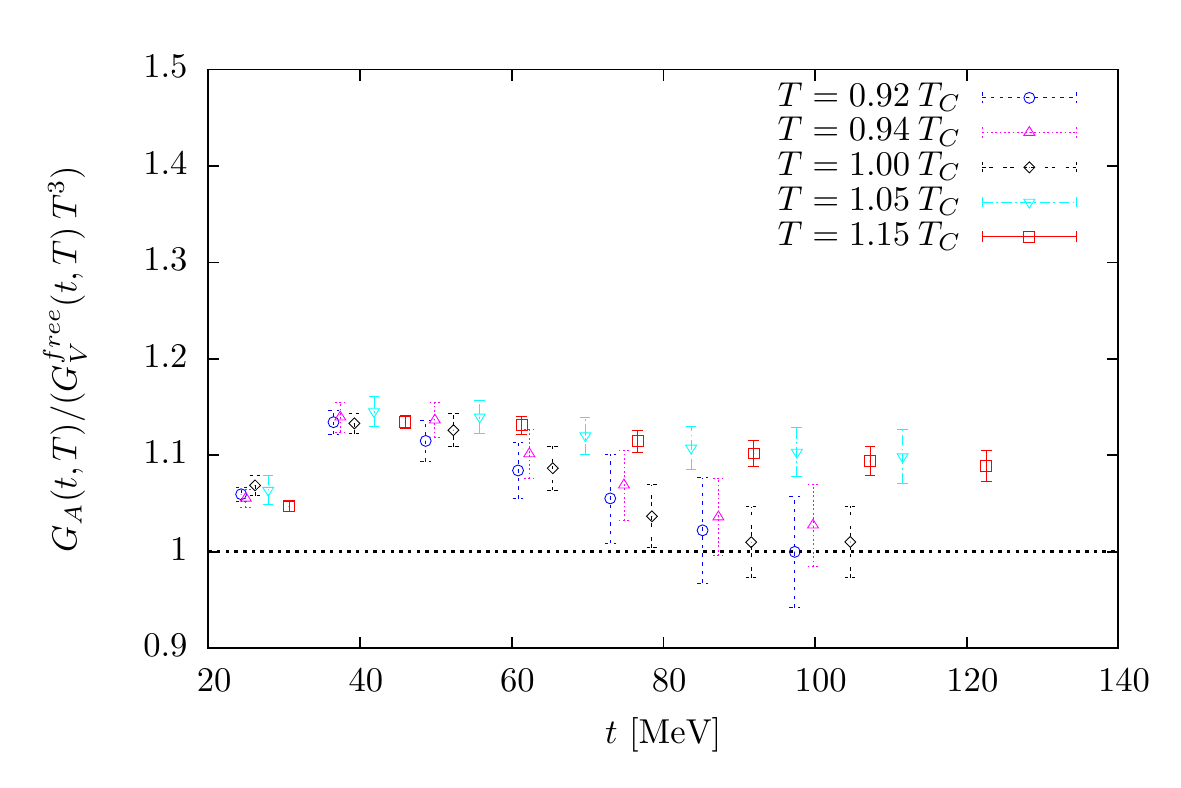}\hspace{0.5cm}\includegraphics[width=6.4truecm,height=4.7cm]{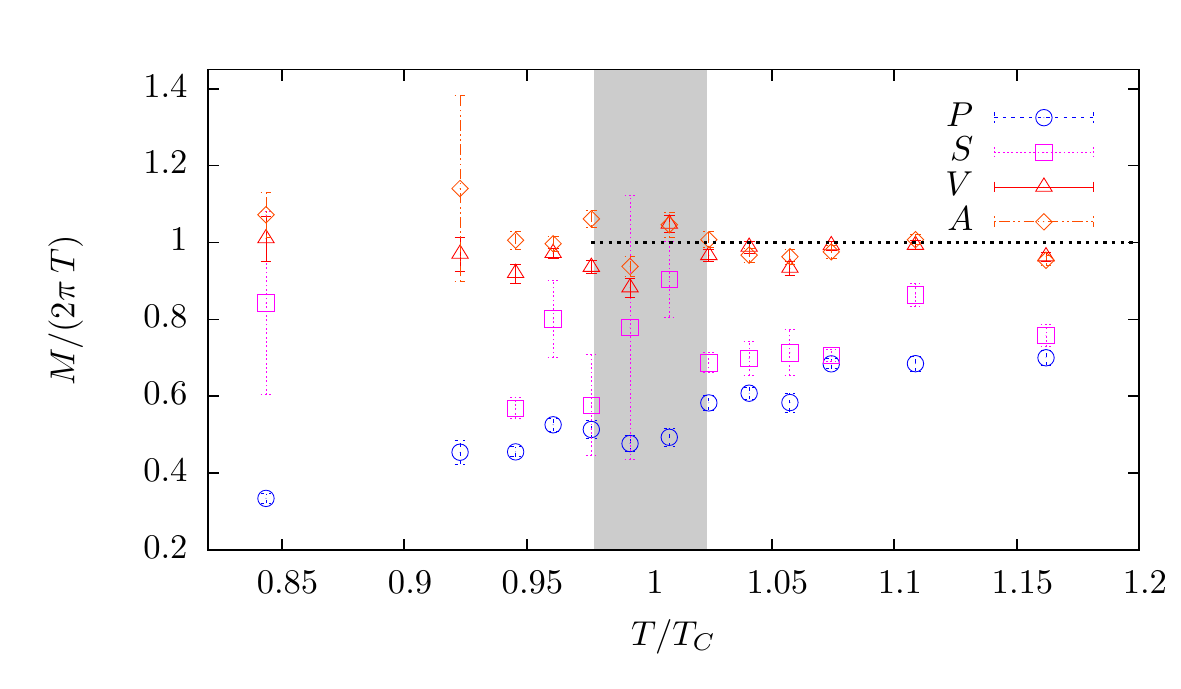}
\label{fig:BB}
\caption{Vector correlators in the plasma, from simulations with two flavors of improved Wilson fermions on large lattices (left), and summary of results (right)(Ref. \cite{Brandt}).}
\end{figure}
\subsection{The universality class for $N_f=2$}
\begin{figure}[b]
\includegraphics[width=6truecm]{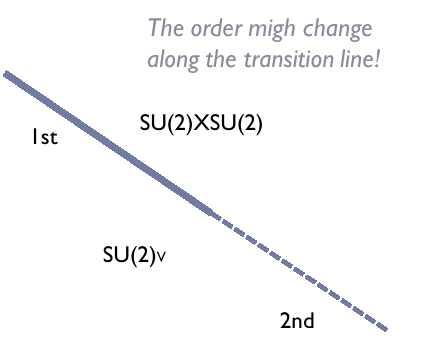}
\hskip 0.5cm
\includegraphics[width=6truecm]{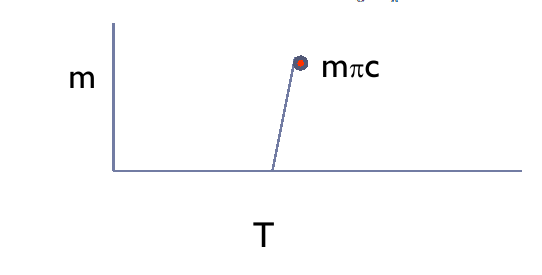}
\caption{Sketchy view of a possible behaviour of the transition line: the
order might change along the line even when the symmetries of the two
phases are the same(left); the region of the first order transition is
characterised by the endpoint at finite mass, in the universality
class of the $Z_2$ Ising model (right).}
\label{fig:order}
\end{figure}
\begin{figure}
\vskip -9 cm
\includegraphics[width=10truecm]{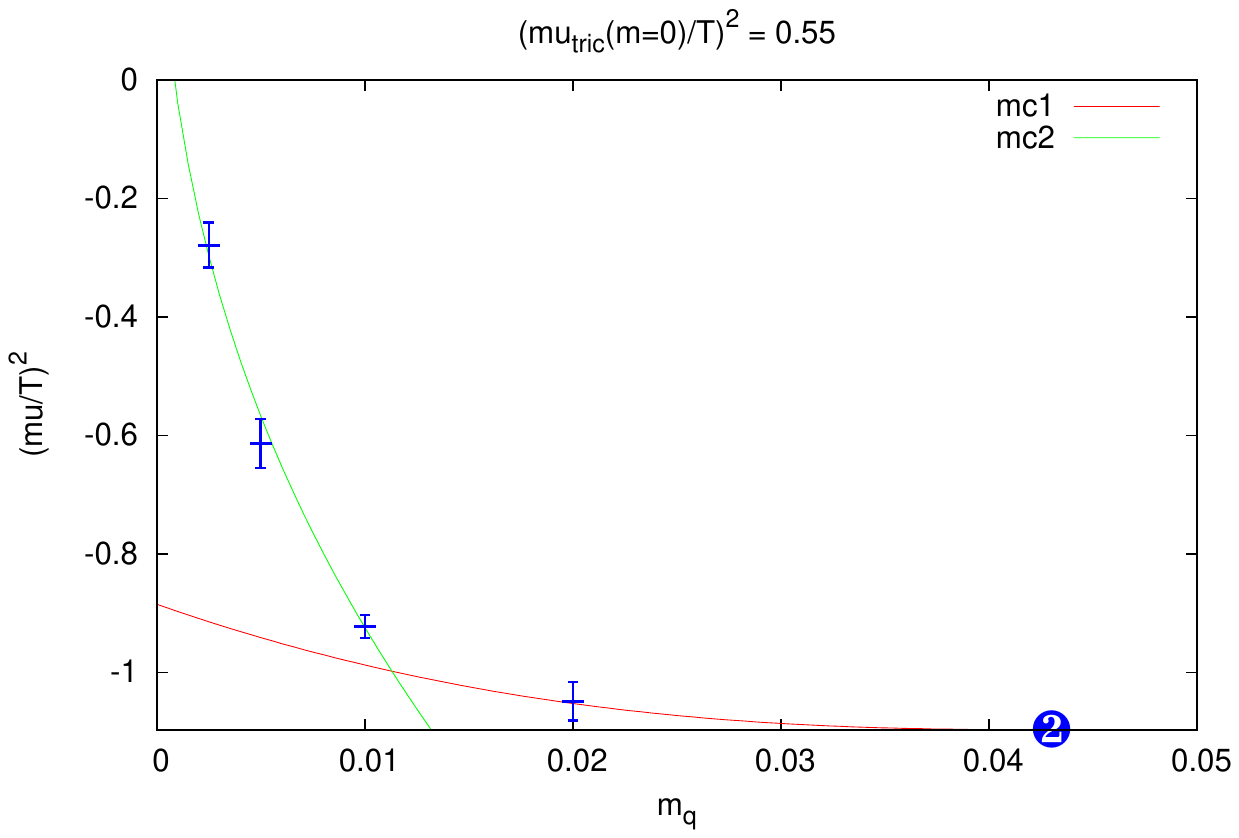}
\hskip -4 cm
\includegraphics[width=10truecm]{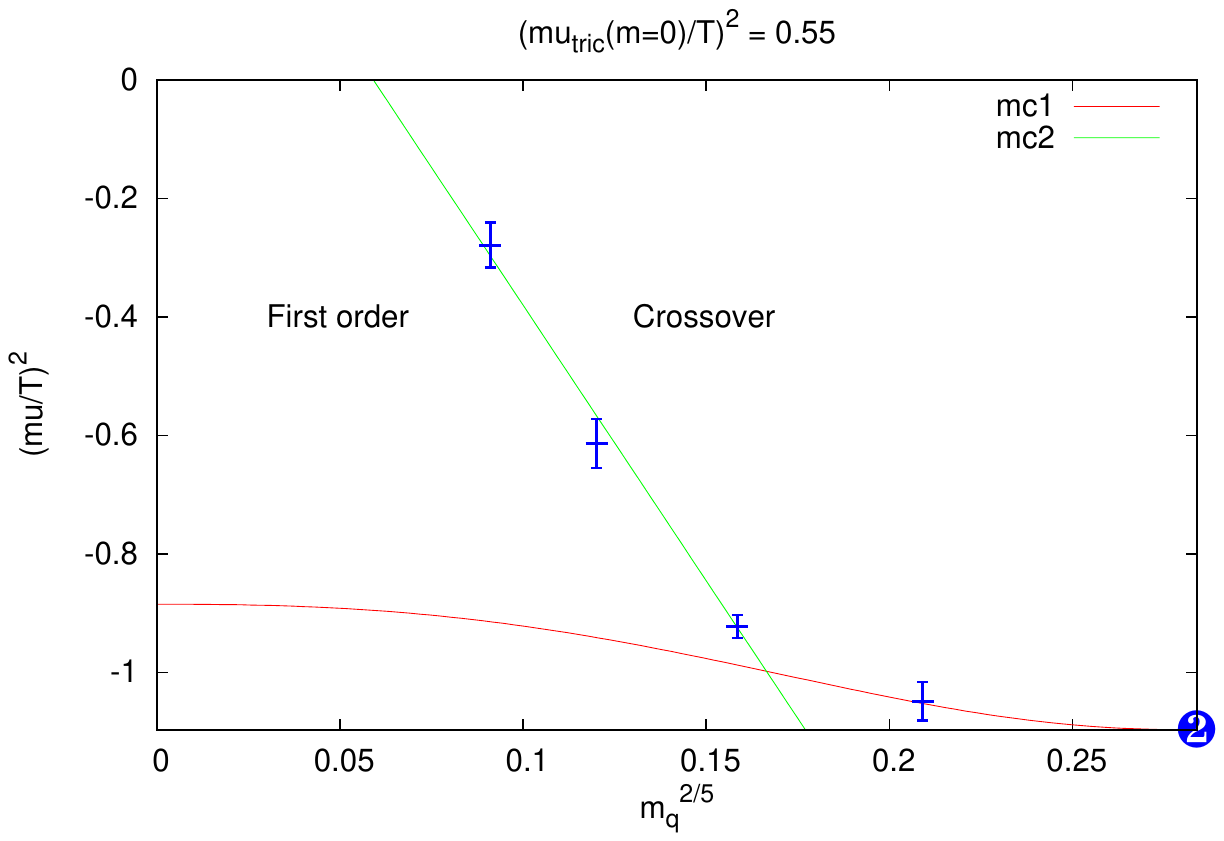}
\caption{Search for universality class of the $N_f=2$ theory : the
position of the endpoint of the first order transition for imaginary
$\mu$ versus the bare mass (the exponent in the right diagram is motivated
by the $Z_2$ universality class associated to the endpoint). The transition becomes of second order
when the endpoint hits zero, and based on these plots the authors
suggest that this might happen for a positive $\mu^2$. The results 
have been obtained for $N_t=4$ unimproved staggered fermions hence
they are far away from the continuum (Ref. \cite{Bonati}).}
\label{fig:genoa}
\end{figure}
As noted above, the order of the transition
for two flavor QCD is a dynamical issue which depends on the magnitude 
of the residual breaking of $U_A(1)$ at the transition point.

The situation can be depicted as in Fig.\ref{fig:order}, left: 
the order might well change along
the transition line even if the symmetries are the 
same on both sides. 
Since the quarks' chemical potentials do not alter the chiral
symmetries in the Lagrangian one can study a possible realisation
of this scenario by  varying $\mu$, away from the condensation phases,
and see whether one can observe first and second order transitions.
\begin{figure}[b]
\centering
\includegraphics[width=12truecm]{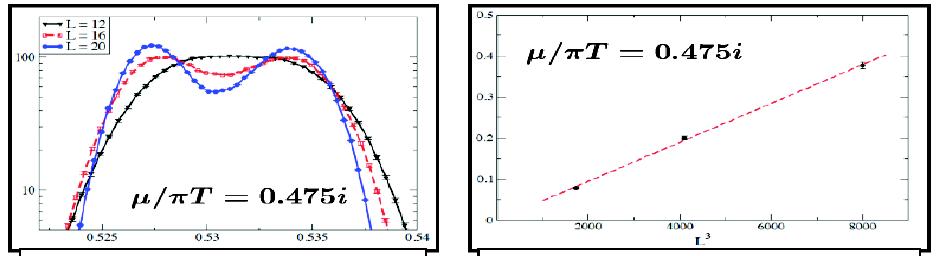}
\caption{Direct observation of a first order transition at non-zero 
imaginary isospin density. If this signal would persist till 
zero chemical potential the $N_f=2$ transition would be of first order
(Ref.\cite{Cosmai}). }
\label{fig:bari}
\end{figure}
\begin{figure}
\begin{minipage}{0.45 \textwidth}
\includegraphics[width=6.4truecm, height=4.5cm]{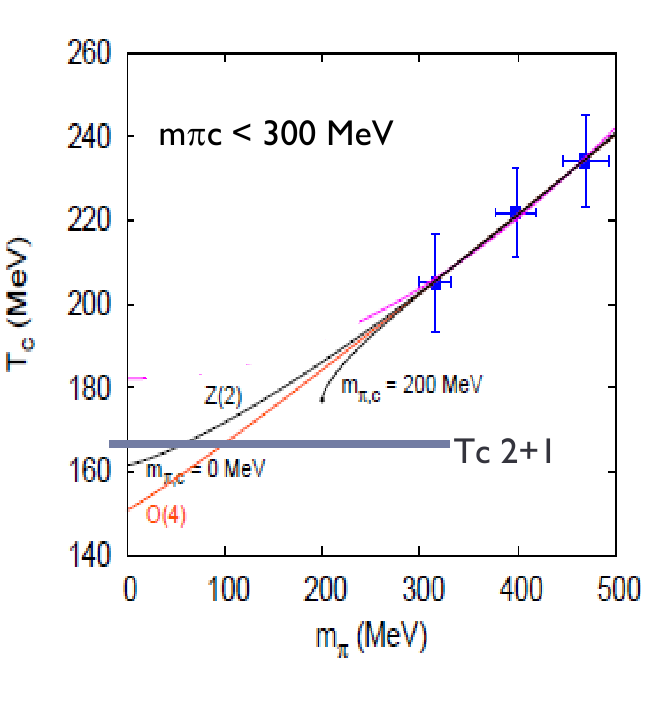}
\caption{Results on the crossover temperature for $N_f=2$, 
with chiral 
extrapolation as motivated by the plausible universality classes 
(Ref. \cite{Burger}).}
\label{fig:florian_tc}
\end{minipage}
\hspace{0.2cm}
\begin{minipage}{0.45 \textwidth}
\includegraphics[width=0.9\textwidth]{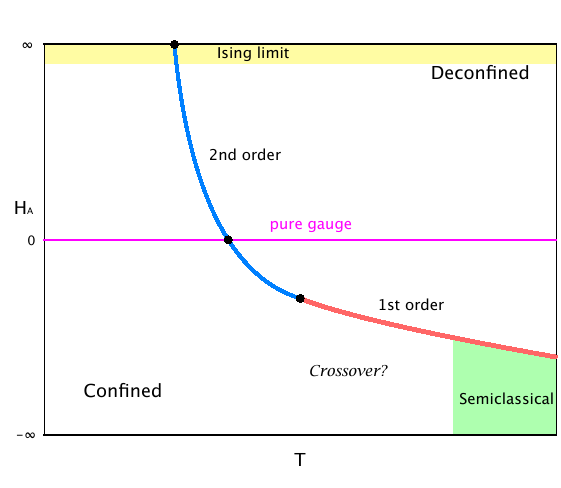}
\caption{The phase diagram in an enlarged parameter space
(Ref. \cite{Ogilvie}).}
\label{fig:moc}
\end{minipage}
\end{figure}
This strategy is pursued \cite{Bonati,Cosmai,Cea:2012ev}  
with imaginary chemical potential.  
Fig.\ref{fig:genoa} shows the results with imaginary
 baryonic $\mu$\cite{Bonati}, and Fig.\ref{fig:bari}
shows the results for imaginary isospin $\mu$\cite{Cosmai}.
The authors
argue that following the fate of the endpoint of the first order transition,
sketched in Fig.\ref{fig:order}, right, 
 is a practical way to decide  the order of the transition at $\mu=0$.
In particular one might want to locate the turning point between
a first and a second order transition: since first order transitions
were observed for a non-zero imaginary chemical potential
for $N_f=2$, a turning
point for real $\mu$ implies that the phase transition at zero
chemical potential is of first order\cite{Bonati,Cosmai}.   
Current extrapolations suggest that this is realised 
for $N_t=4$, for unimproved staggered fermions, hence
far from the continuum limit. 

The transition for two flavors close to the continuum limit is being 
investigated with Wilson fermions\cite{Burger},
and some results are shown in Fig.\ref{fig:florian_tc}. The plot depicts
the current situation with Wilson fermions, maximal twist and improved
gauge sector. 
The extrapolation to the chiral limit can be carried out either with
a second order ansatz, or assuming a critical endpoint smaller 
than $300$ MeV.
The data do not discriminate between these two possibilities.
This implies that the order of the $N_f=2$ 
transition in the massless continuum limit is still undecided, and 
suggests that the endpoint of the first
order transition at $\mu=0$, if any, is for $m_\pi < 300$ MeV.

\section{The transition region - confinement, gauge dynamics, external fields,
$\theta$ term}
The discussions in the previous Section focused
on the fermionic sector of the theory,
and the associated chiral symmetries.
The ''obligatory'' step at this point is to remind ourselves that 
gauge dynamics and confinement are crucial, yet mysterious,
aspects of the transition, in particular due to the lack of a bona-fide 
order parameter for deconfinement.

Abelian gauge theories are an important exception :
in these theories  the confinement transition is understood as a 
condensation 
of topological defects -- in the case of  four dimensional $U(1)$, 
for instance, strings of magnetic monopoles \cite{Polyakov}. 

Ogilvie proposes to search for a continuous path between non--abelian theories
and abelian ones by suitably enlarging the parameter space\cite{Unsal:2008ch,Ogilvie}.
The considered phase diagram is depicted in Fig.\ref{fig:moc}. By supplementing
the Yang-Mills $SU(2)$ action with a double trace term, one can show that  
the confined phase of pure $SU(2)$ and the high temperature 
confined phase of the extended model are continuously connected;
the latter is amenable to a semiclassical treatment and in turn
is continuously connected to the confined phase of the abelian model.

\begin{figure}
\centering
\includegraphics[width=6truecm]{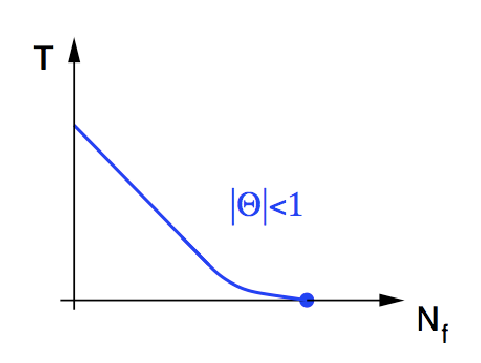}\includegraphics[width=5truecm]{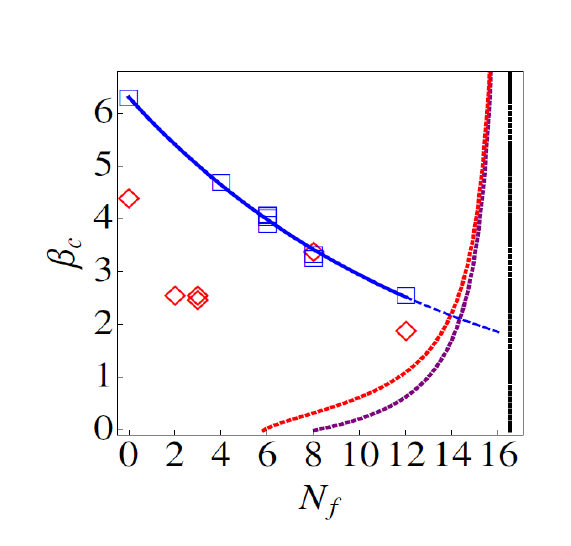}
\caption{The transition as a function of matter content (left) and the
associated coupling according. Results from various lattice
groups (Ref. \cite{Liao:2012tw}).}
\label{fig:mag}
\end{figure}

A more direct interpretation of the transition in terms of magnetic 
monopoles has been recently proposed by Shuryak\cite{Liao:2012tw}.
In Fig.\ref{fig:mag} lattice data for different 
$N_f$ are used to estimate the strength of the coupling at the transition,
which is then compared with the coupling at the IRFP\cite{Schaich,Nunes}. 
In particular the data marked with a diamond were obtained with the same
action, and different number of flavors, and can be meaningfully considered
together\cite{Miura:2011mc}. The coupling at the transition becomes stronger
and stronger with increasing $N_f$:  
the proposed explanation  is that light fermions 
can occupy the  (chromo)magnetic monopoles, making them unsuitable for 
Bose Einstein  condensation.  Lattice simulations measuring magnetic
monopole properties can confirm this proposal, which, if true, would lend
support to confinement scenarios based on monopole condensation in QCD. 
A further related observation is the emergence of the conformal window
of QCD for large number of flavours -- 
 reviewed by Joel Giedt\cite{Giedt} -- as the zero temperature limit
of a super--strongly interactive Quark Gluon Plasma.

\begin{figure}
\centering
\includegraphics[width=6 truecm,height=4.5cm]{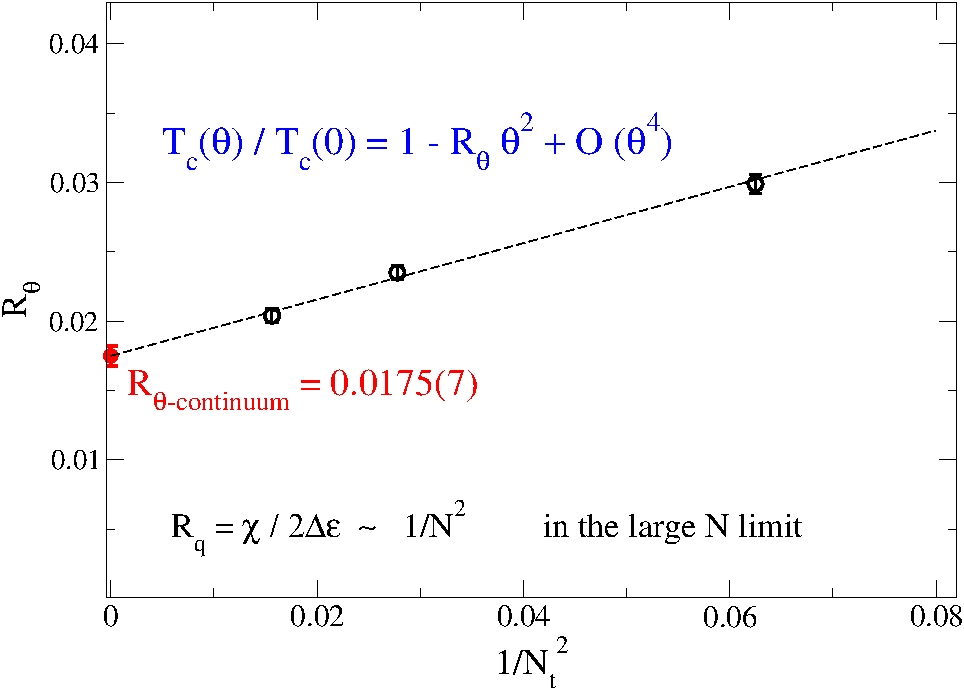}\hspace{0.9cm} \includegraphics[width=6truecm,height=4.9cm]{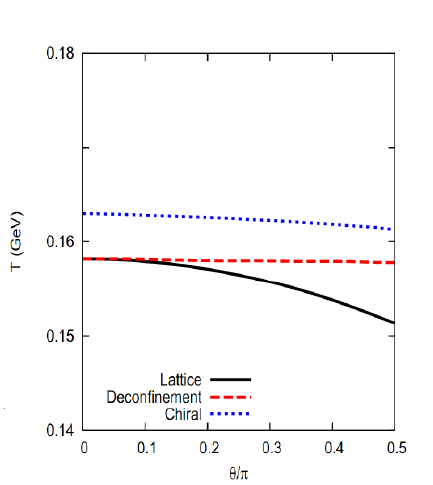}
\caption{News on the $\theta$ vacuum: the continuum extrapolation of
the slope of the critical temperature from lattice simulations at imaginary
$\theta$(left, Ref. \cite{Negro}), and from model studies(right, Ref. \cite{Sasaki}).}
\label{fig:theta}
\end{figure}

It has been proposed that fluctuations of topological charge
might play a role at the transition, producing measurable
$P$ and $CP$ violations in heavy ion collisions. In the
QCD Lagrangian $P$ and $CP$ violating terms enter via the $\theta$
term $\theta Q(x)$, $Q(x)$ being the topological charge density.
Despite the experimental bound  $\theta < 10^{-10}$ it is
still conceivable that such term affects the critical temperature
\cite{Unsal:2012zj,Poppitz:2012nz}.
New studies on the $\theta$-dependence of the deconfinement
temperature have been presented\cite{Negro,D'Elia:2012vv,Sasaki}, 
the former numerical and the latter based on a model analysis. 
In both cases $T(\theta)$ was parametrised as 
$T(\theta) = T_c - R_\theta \theta^2$, and  
the coefficient of the
quadratic term $R_\theta$  has been estimated. The
results are shown in Fig.\ref{fig:theta}. 
Both studies agree on  the sign of $R_\theta$, as well 
as on its very little magnitude, with a substantial agreement
within the largish systematic errors.

Finally, an interesting line of research  is the effect of an external 
field on the phase transition\cite{Kharzeev:2012ph}. 
As magnetic fields as large as 
$10^{14}$ Tesla can be produced at the QCD transition a lattice 
effort has started to compute the effects of strong magnetic fields 
on the QCD transition, and on the magnetic catalysis --
a possible enhancement of chiral symmetry breaking due to
the effects of the external fields. Although intuitively an external
magnetic field should favour chiral breaking,
models' analysis gives non-univocal results, and, again, lattice
calculations are mandatory to settle the issue.  
 Details of the dynamics are still
under discussion, particularly close to the chiral limit
and I refer the interested reader to a recent 
review for details\cite{D'Elia:2012tr}. Recent
results have been obtained   for two \cite{Ilgenfritz:2012fw} and  
three colors\cite{Bali:2012zg}. 

\section{Fluctuations}
Lattice QCD simulations are being carried out in a GranCanonical ensemble.
This allows fluctuations of conserved quantities -- baryon number,
electric charge, strangeness -- for instance, the baryon
number should be zero on the average, but it can fluctuate configuration by
configuration. In ordinary conditions
 fluctuations are  quite small: for
instance one can compute the distribution of baryon number in ChPT 
and find that it is 
described by a narrow Gaussian
\cite{Lombardo:2009rt,Lombardo:2009aw} 
while when the temperature increases the  
distribution broadens and might change 
shape\cite{Lombardo:2009rt,Lombardo:2009aw}.
The fluctuations
thus reflect  properties of the thermal
medium and of the critical point\cite{Stephanov:2008qz}. 
 
In general, a synthetic way to describe a distribution is via its momenta
\begin{equation}
\mu_r = <X^r>  = \int_{-\infty}^{+\infty} X^r f(X) dX
\end{equation}
The momentum generating function 
$M(\xi)  = <exp (\xi X) > = \sum_{r=0}^\infty \frac {\xi^r}{r!} <X^r> $ 
is a simple way to combine momenta into a single expression and we 
recognise that the
momenta are linked with the coefficients of the Taylor expansion of the 
generating functional
$<X^r> = [\frac {d^r}{d \xi^r}  M(\xi)]_{\xi =0}$.
Consider then the pressure 
$p = T/V ln {\cal Z}(T, \mu)$ and its Taylor 
expansion in powers of $\mu/T$ around $\mu=0$:
\begin{equation}
\frac {p(\mu/T)}{T^4} = \sum_{n=0}^\infty c_n(T) (\frac{\mu}{T})^n
\end{equation}  
and we recognise that the zero density
fluctuations of the quark number
density $n(\mu) = \frac {\partial p(T,\mu)}{\partial \mu} $
are  given by the coefficients $c_n(T)$. Similar reasoning
holds for the other conserved charges. 
Results for the momenta have been successfully compared with Hadron Resonance
Gas till temperatures very close to $T_c$
\cite{Bazavov:2012jq,Borsanyi:2011sw}.

\begin{figure}
\centering
\includegraphics[width = 0.7\textwidth,height=6cm]{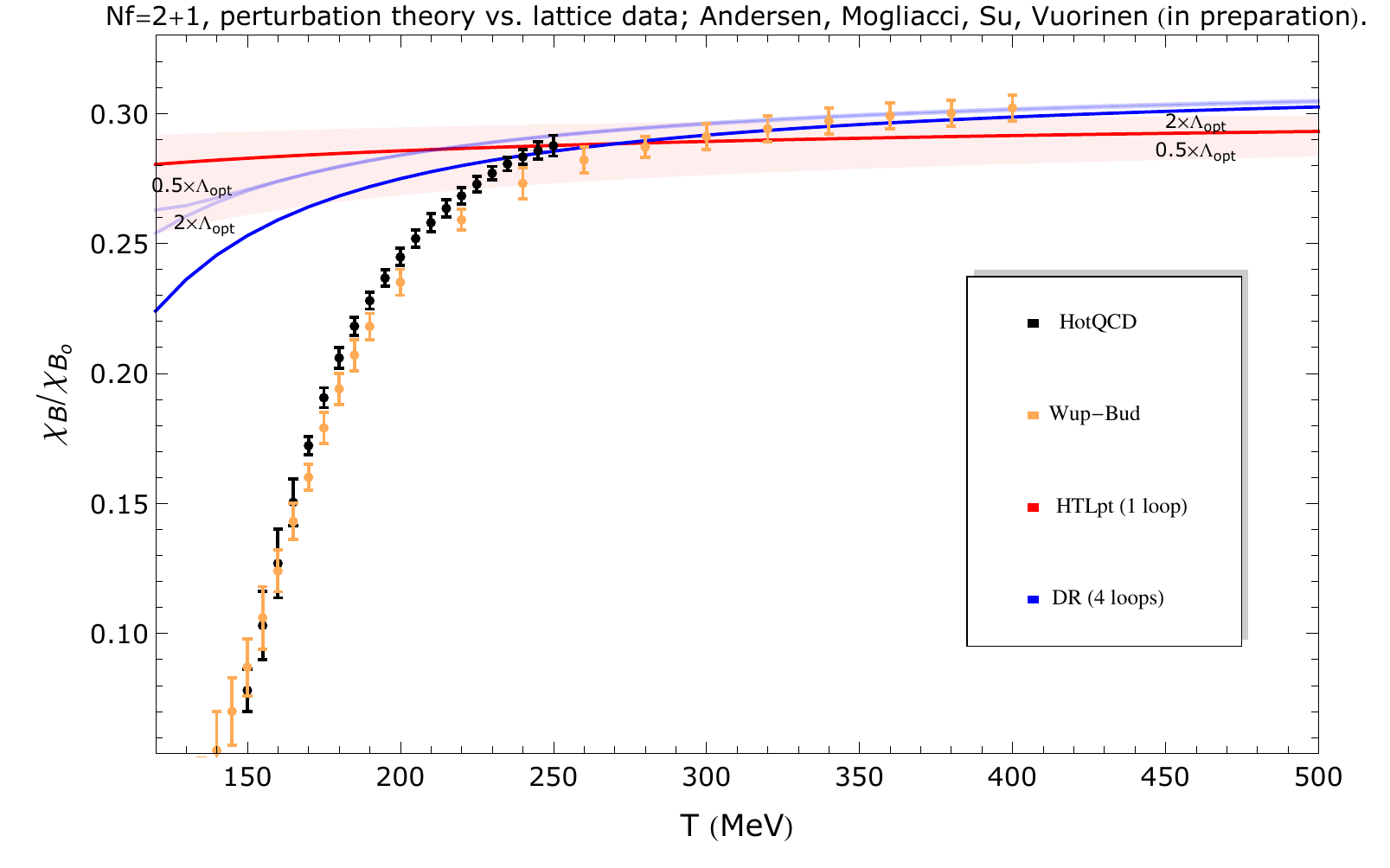}
\caption{Lattice results for the baryonic susceptibility vis-a-vis analytic
studies. Note the agreement between different lattice calculations, and
with the 4 loops dimensionally reduced theory (Ref. \cite{Vuorinen}).}
\label{fig:aleksi}
\end{figure}

At the time of the meeting new analytic results 
from A. Vuorinen and collaborators \cite{Vuorinen} became available, 
followed by a publication \cite{Andersen:2012wr}. 
The excellent agreement between analytic and numerical
results -- which can be appreciated in 
Fig.\ref{fig:aleksi} -- is in part due to the elimination of the
pure gauge component (by taking the derivatives w.r.t to $\mu$.) 
Note also in the same figure the very
good agreement between the results of the Wuppertal-Budapest 
collaboration and 
HotQCD:  these measurements have been performed 
on the same set of gauge fields discussed above 
and the agreement here again corroborates the optimistic view that the 
residual minor disagreements are almost solved. 

\begin{figure}[b]
\centering
\includegraphics[width=6truecm,height=6cm]{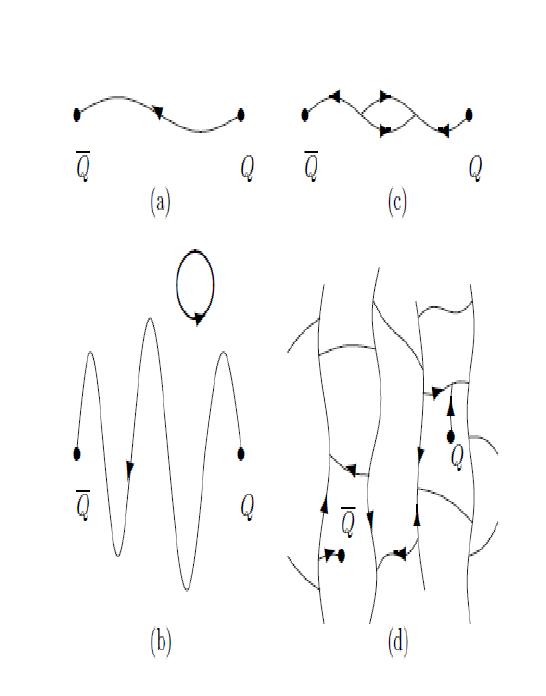}
\hspace{1cm}\includegraphics[width=8truecm,height=6cm]{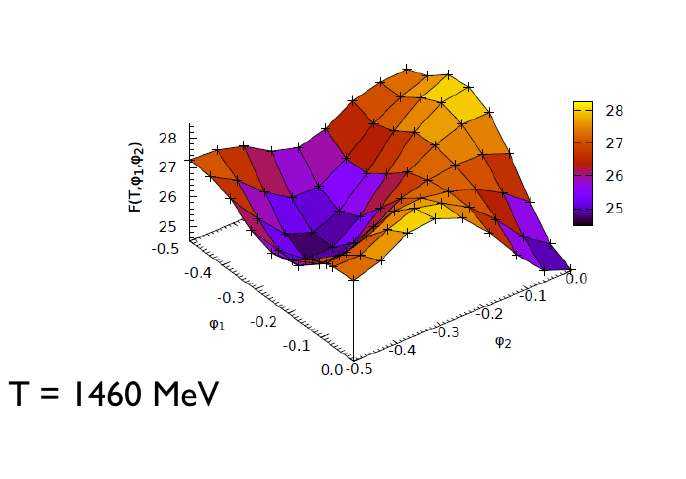}
\caption{Describing fluctuations: the flux tube model by Patel
(left, Ref. \cite{Patel}). The Polyakov loop analysis by Gattringer (right,
Ref. \cite{Diakonov:2012dx}).}
\label{fig:patel_gattringer}
\end{figure}

A completely different approach to the dynamics
of fluctuations is offered by the 
flux tube model presented by Patel\cite{Patel}:  the idea is that
the 
fluctuations should be 'visible' in the gauge fields, and one
can conceive, within the flux tube models, 
one way to visualise the gauge field dynamics
in position space. The 'rules of the game' 
in a first approximation are simple:
quarks and anti-quarks are bond by a string in a $\bar Q Q$ state. 
The string can break via the formation of a light-heavy meson, 
$\bar Q Q \to q \bar Q + \bar q Q$ , or in the plasma as
 $\bar Q Q \to \bar Qq + qq Q$, see Fig \ref{fig:patel_gattringer}.
In this way one might hope to study 
multi-correlation particles which are useful for phenomenology,
and highlight the microscopic structure of the medium, for instance
by distinguishing a dilute liquid from a cristal-like structure.  

\section{Quark Gluon Plasma - a new world}
The physics of the Quark Gluon Plasma is of course addressed by 
most of the studies described before. Here -somewhat arbitrarily -  
I collect those studies which are motivated by the
desire of a deeper and general 
theoretical understanding of the physics of this
unusual state, without reference to specific models. 

\begin{figure}
\centering
\includegraphics[width=7truecm,height=4.5cm]{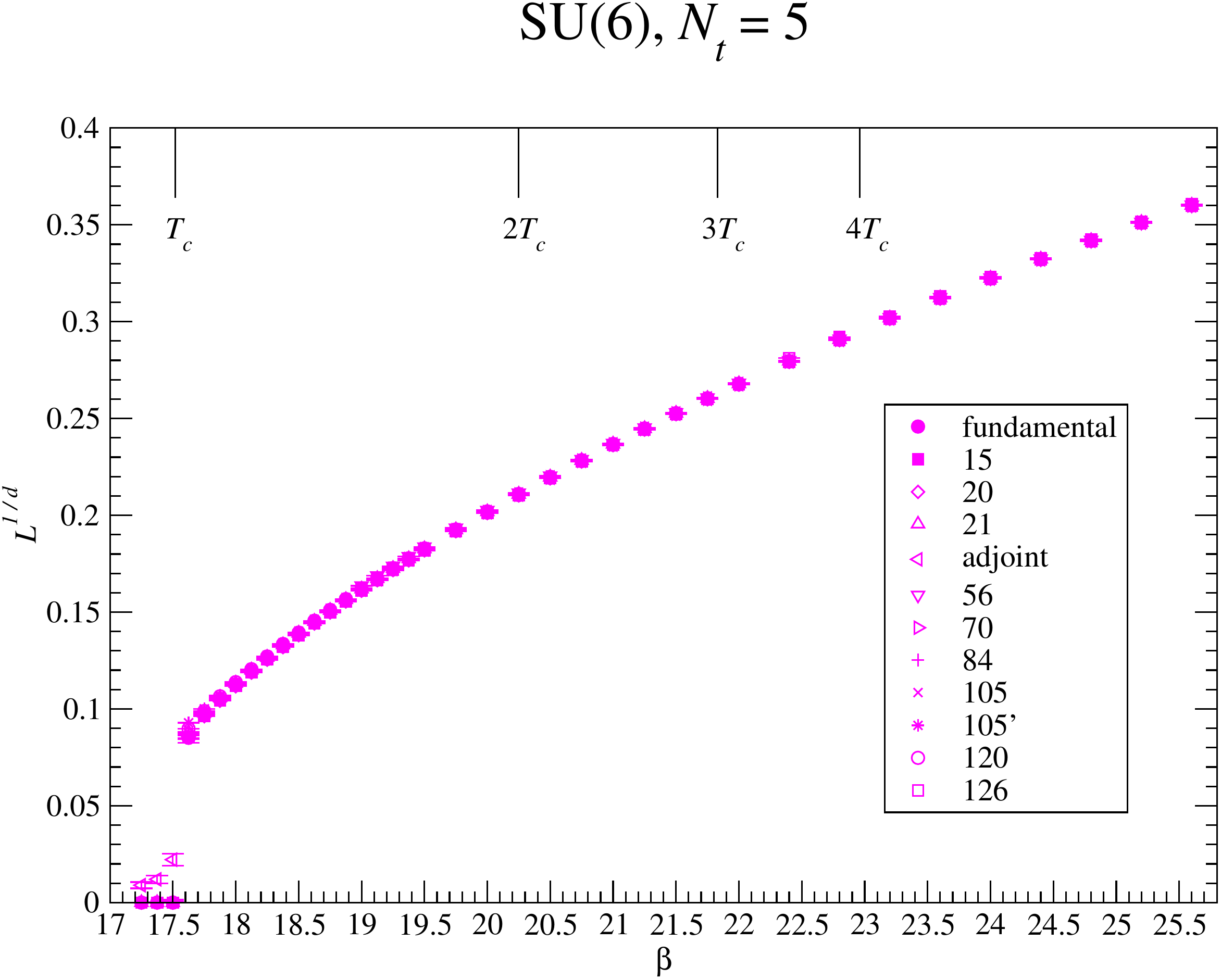}\hspace{1cm}
\includegraphics[width=7truecm,height=4.5cm]{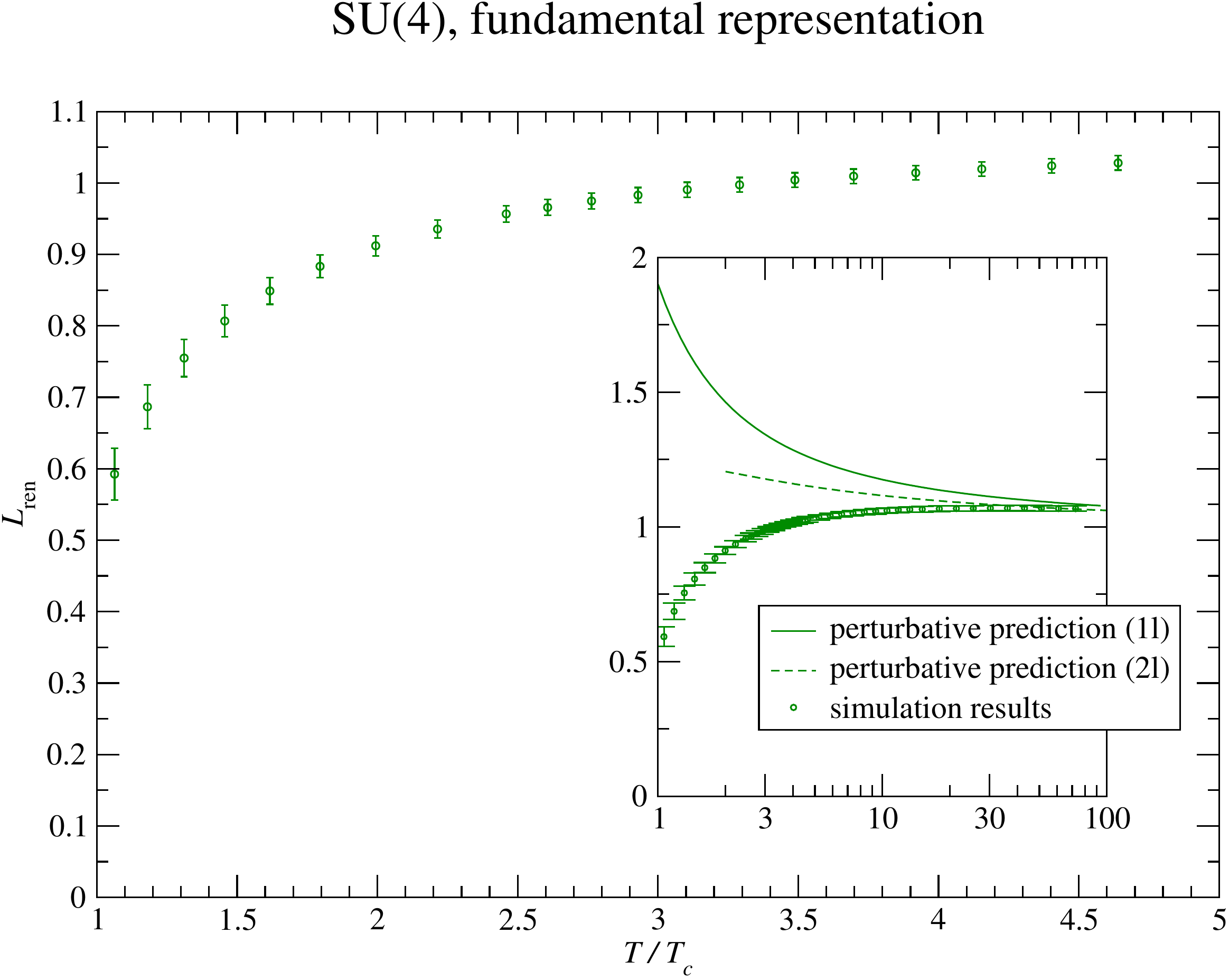}
\caption{The left diagram shows 
the bare loop for $SU(6)$, and different representations, 
rescaled assuming Casimir scaling--the universality of the results
support the assumption. The right-hand 
diagram  shows the same renormalised loop in the fundamental representation for
$SU(4)$--in both cases the behaviour is very similar to the one 
observed in $SU(3)$ (Ref. \cite {Panero}).}
\label{fig:panero}
\end{figure}

One popular way to learn about general features is to enlarge the parameter
space of interest: a large number of colors, or a large number of flavors.
For a large number of colors $N_c$ 
--  reviewed by M. Panero\cite{Panero} -
we have one very interesting finite temperature 
result\cite{Mykkanen:2012ri}. Panero and
collaborators have studied the behaviour of the Polyakov loop in the 
deconfined phase of a $SU(N_c)$ Yang-Mills theory. 
Different representations
and different $N_c$ exhibit a near universal behaviour once geometric 
scaling factors
are taken into account, see Fig. \ref{fig:panero}. 
This observation - so far restricted to
pure gauge systems - lends support to large $N_c$ approaches to
the analysis of the high temperature phase of QCD. 

It is believed that as  chiral symmetry is no longer
spontaneously broken, the only remaining scale in the massless model
is the temperature. Would the eigenvalue spectrum of QCD give information
on any remaining scale? Kovacs\cite{Kovacs} 
has shown that this is indeed the case - 
the equivalent of the Thouless energy was found to scale as 
the temperature itself.

Attempts at modelling the Quark Gluon Plasma by use of Polyakov 
loop dynamics
has produced the observation that $Z_3$ metastable states might become 
important at $T > 750$ MeV \cite{Deka:2010bc}. 
Such metastable states with dynamical fermions
might have an influence on bubble formation and 
on the system's dynamics in general.

How to detect directly the dynamics of the Polyakov loop? 
Diakonov, Gattringer 
and Schadler \cite{Diakonov:2012dx} propose 
a free energy for the untraced Polyakov loop $X$ 
\begin{equation}
F(T,X) = T \int_0^\beta d \beta ' <S>_{T,X,\beta'}
\end{equation}
where $X$ is defined by
\begin{equation}
P = \frac{1}{V_3} \sum_{t=1}^{N_t} U_4(\vec x, t) = Tr X
\end{equation}
and
\begin{equation}
<S>_{T,X,\beta'} = - \frac{\partial}{\partial \beta '} \ln Z(T,X,\beta')
\end{equation}
This definition, so far tested in the pure Yang Mills case, see Fig. 21, right,
compares  very well with high $T$ perturbation theory and opens 
the way to a rigorous 
approach to the study of Polyakov loop dynamics. 

\section{Quark Gluon Plasma confronts experiments}
How far we can go with comparing our results with experiments?
For orientation, we know that during 
Au--Au collisions at 200 GeV at 
RHIC the system reached  up temperatures in the range 340--380 MeV, the current
LHC runs -- 2.76 TeV -- explored up to 420--480 MeV 
with possibly hot spots of about 500--600 MeV. 
LHC at maximum energy will reach about 1GeV - six times $T_c$
\cite{Heinz}. 
\begin{figure}[b]
\centering\includegraphics[width=0.7\textwidth,height=7cm]{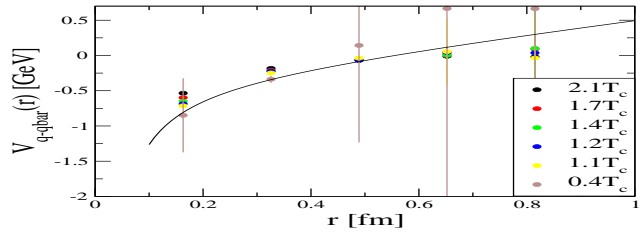}
\label{fig:chris}
\vskip -4cm
 \caption{A new strategy for the potential analysis: Interquark potential obtained by reverse engineering the Schroedinger
equation after calculations of the wavefunction (Ref. \cite{Allton}).}
\end{figure}
The recent CMS results\cite{:2012fr} 
are a good starting point for the discussion: 
the invariant mass plot measured in pp shows the expected three bumps in 
correspondence to
the fundamental $\Upsilon$  
state and the first two radial excitations. 
When the same distribution is being
measured by colliding lead ions, the lowest peak is almost stable, 
while the two higher peaks have almost disappeared.
A natural thing to do is to compare these observations with the heavy quark 
spectrum computed on the lattice - a notoriously  difficult task. 
see eg. Ref\cite{Kaczmarek:2012ne} for a recent comprehensive overview.

In early times a lattice computed potential was used as input to the 
Schroedinger equation for heavy mesons. Later on, issues were risen concerning 
the appropriate potential to use, and most important
it was finally realised that the correct potential should include an imaginary
component to account for the broadening of the 
state\cite{Brambilla:2008cx,Rothkopf:2011db}. 
Alternative calculations of the
potential using the wave function as an input to the Schroedinger equation, 
and a reverse 
engineering to extract the potential itself have been presented at this meeting as well\cite{Allton}, and the results can be seen in Fig. 23.

At the other extreme - w.r.t. quantum mechanical, non-relativistic studies -
we have field theoretic, relativistic calculations of the 
spectral functions:the euclidean charmonium correlator 
is computed\cite{Ding:2012sp}, see Fig. 24 -- in this case on quenched configurations --, 
and the spectral function is then reconstructed by use of by now standard
Maximum Entropy Method analysis (MEM)\cite{Asakawa:2000tr}. 
\begin{figure}
\centering
\includegraphics[width=14truecm,height=6.cm]{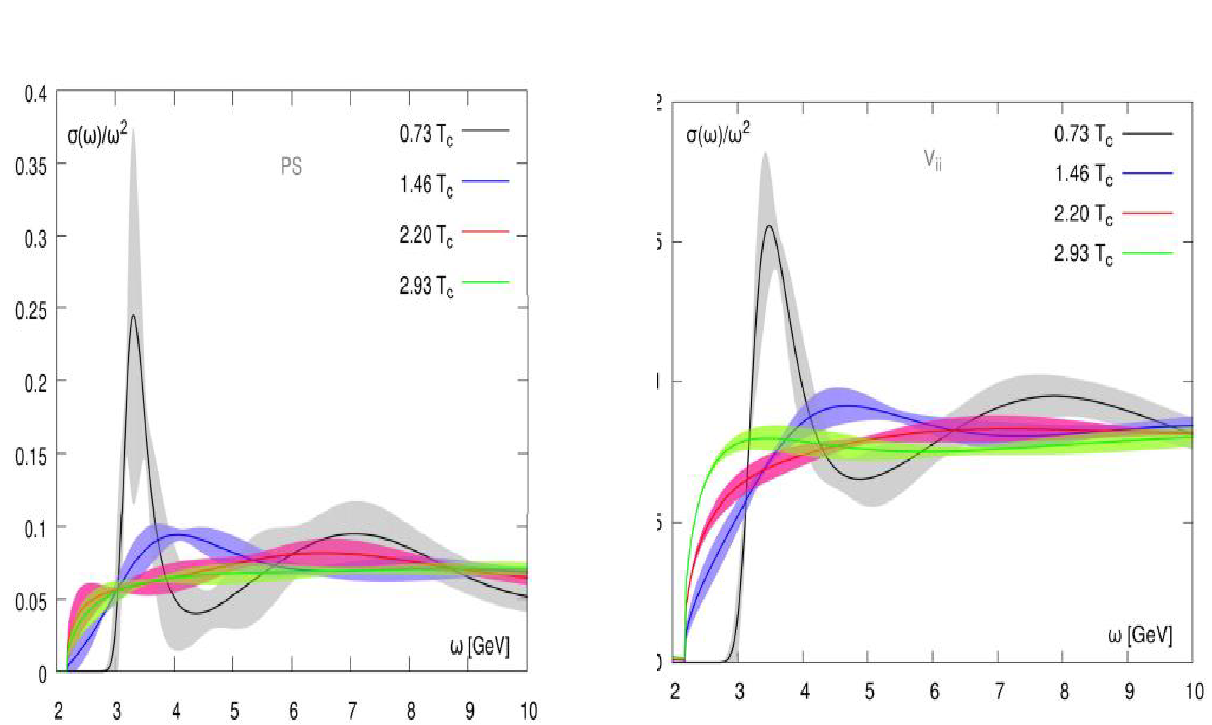}
\label{fig:charmonium}
\caption{Charmonium spectral functions from a full relativistic 
calculation (Ref. \cite{Ding:2012sp}). }
\end{figure}

All these approaches give results broadly 
consistent with each other, and with intuition.
However a solid understanding of the systematics is still lacking.

An improvemed implementation  of the Maximum Entropy
Method  has been presented by A. Rothkopf \cite{Rothkopf}.
The method proposes to use an extended search space, and preliminary
successful results have been obtained by use of mock data: the
numerical behaviour is more stable, and the resolution is satisfactory.

One further approach consists of evaluating NRQCD propagators over 
dynamical gauge fields: the reconstruction of the 
spectral functions is made easier\cite{Aarts:2011sm} 
as in this case  the 
spectral relation  reduces to
 \begin{equation}
 \label{eq:Gnr}
 G(\tau) =
 \int_{-2M}^\infty\frac{d\omega'}{\pi}\, \exp(-\omega'\tau) \rho(\omega')
 \;\;\;\;\;\;, 
 \end{equation} 
 As a result, the spectral function is
merely the inverse Laplace transform of the correlator,
and all problems associated with
thermal boundary conditions are absent. Remarkably, the 
NRQCD bottomonium spectral functions\cite{Aarts:2011sm} show 
a suppression of excited states 
consistent with the CMS results \cite{:2012fr}.
The new NRQCD results\cite{Aarts:2012ka}  presented this
year include a systematic study of momentum dependence, 
as well as a study of the mass dependence\cite{Kim}.
A successful comparison with the predictions 
of effective models has started, 
together with a more detailed analysis of the systematics.
\begin{figure}
\centering
\vskip-3cm
\epsfig{figure=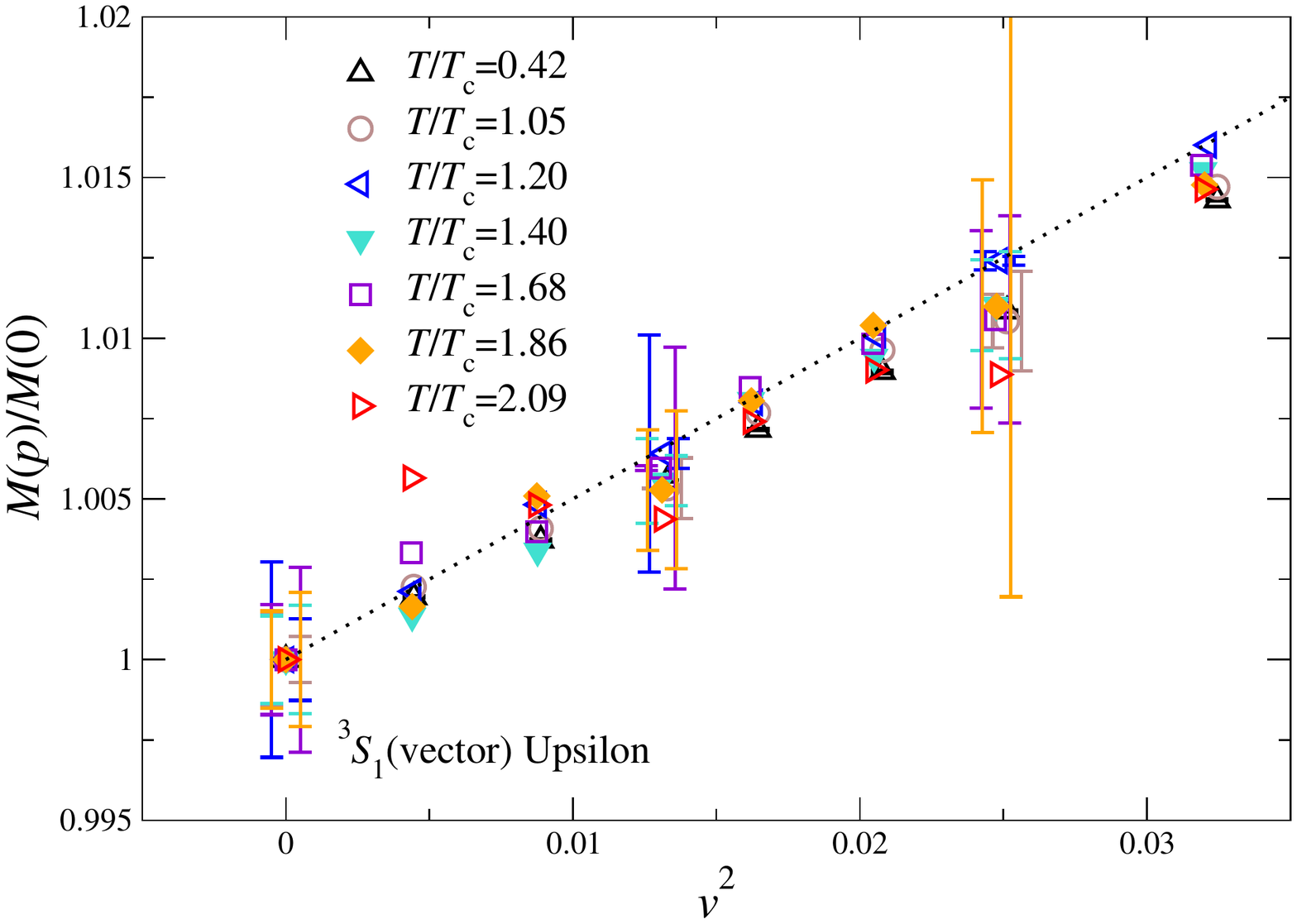,width=0.46\textwidth}
\hspace{0.5cm}
\epsfig{figure=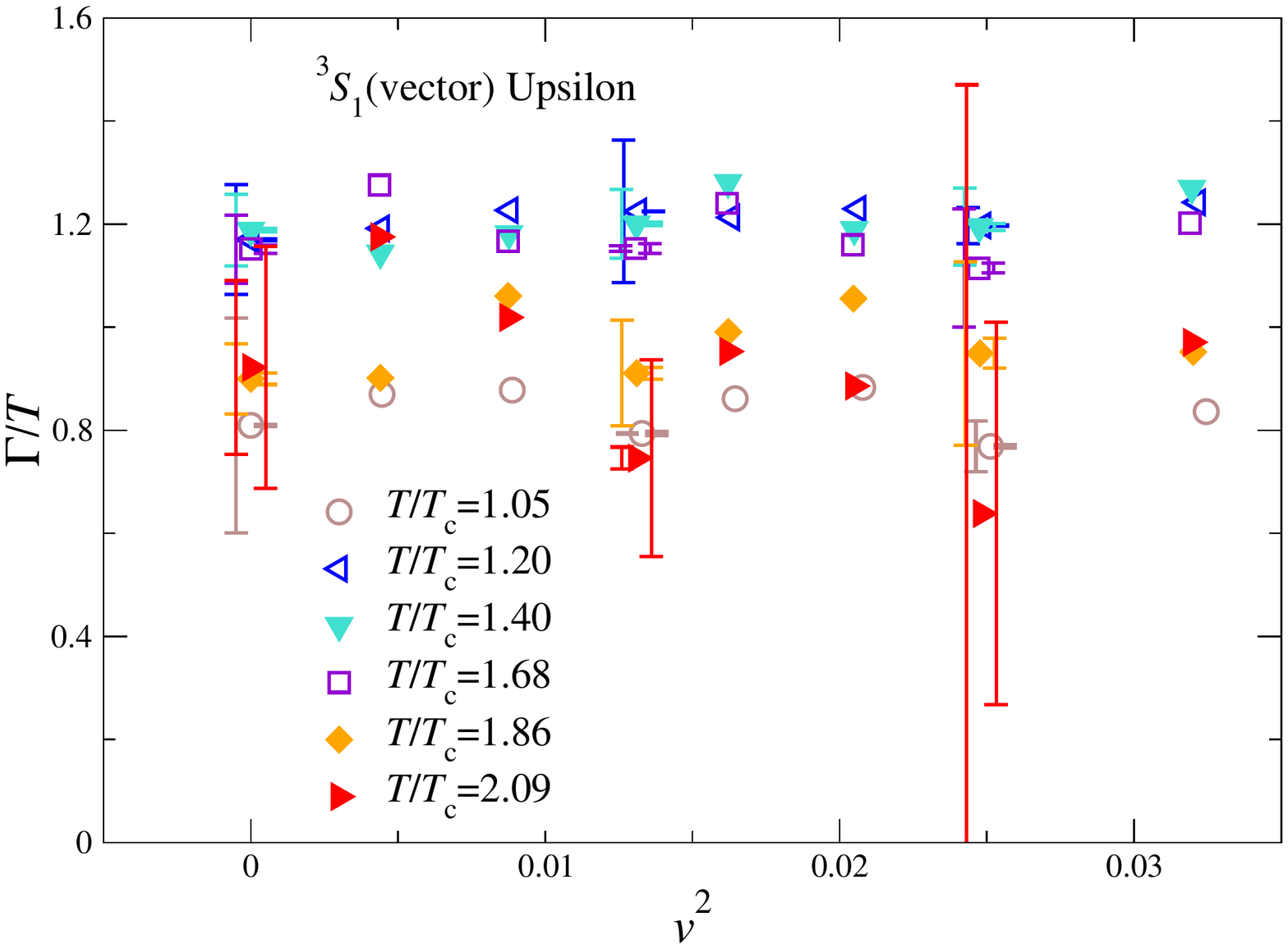,width=0.46\textwidth}
 \caption{Momentum dependence of spectral functions in the plasma,
as seen in the mass and width. 
Position of the ground state peak $M(_{pv})/M(0)$ (left)
and  the upper limit on the width of the ground state peak, normalized
by the temperature, 
$\Gamma/T$ (right), as a function of the velocity squared ($v^2$) in the vector ($\Upsilon$) channel (Ref. \cite{Aarts:2012ka}).}
 \label{fig:mass-width}
\end{figure}

Directly related with the computation of spectral functions are the 
transport coefficients, see e.g Ref.\cite{Adams:2012th} for a 
comprehensive review. 
New results appeared this year concern the  heavy quark diffusion. 
In the framework of a Gaussian and Markovian  Langevin
dynamics,  the diffusion constant $D$  describes the mean square
displacement for times $t$ larger that the relaxation time
 $< |\Delta (x)|^2> = 6Dt $.   In the strong coupling limit
of the SUSY Yang-Mills plasma one finds, using holographic duality
\begin{equation}
D = \frac{2}{\pi T} \frac {1}{\sqrt \lambda}
\label{eq:D}
\end{equation}
Gubser\cite{Gubser:2006qh} suggests $\lambda \simeq 6 \pi$ in QCD,
concluding $D \simeq 1/(2 \pi T)$.  
In a relativistic setup, transport coefficients  are defined as  
\begin{equation}
 \lim_{\omega \to 0} \frac {\rho(\omega)}{\omega^2}
\label{eq:tr}
\end{equation}
where $\rho(\omega)$ is a spectral function in the appropriate channel\cite{Meyer:2011gj}.
For instance $D$  has been extracted from the charmonium 
spectral functions described above\cite{Ding:2012sp},
with results  remarkably close to the one obtained
in strongly coupled SUSY, eq. \ref{eq:D}. 

The diffusion coefficient $D$ can also be extracted from
pure gauge correlators\cite{Banerjee:2011ra}: lattice results
for this have been presented by Datta\cite{Datta}. 
Again the result is close to the one of eq.\ref{eq:D},
hence in remarkable agreement with those from 
Ref.\cite{Ding:2012sp}.

\begin{figure}
\centering
\includegraphics[width=10truecm,height=5.cm]{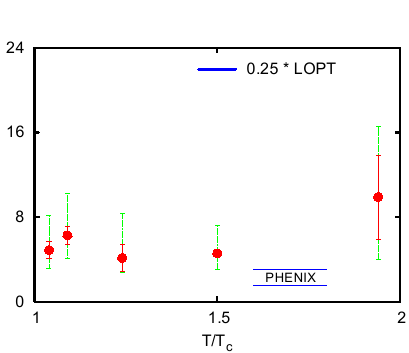}
\caption{Heavy quarks diffusion coefficients : the results,
in impressive agreement with
experimental ones, corroborate a strongly interactive quark gluon
plasma -- the perturbative high T limit  is shown as a short segment on the
right (Ref. \cite{Datta}).}
\label{fig:D}
\end{figure}

The comparison with experimental results  
is shown in Fig.\ref{fig:D}. Needless to say,
it will be of a great interest to see the new 
results from LHC, at higher energies, 
when the plasma is expected to
be less bound, and the transport coefficient should
correspondingly increase. 

To appreciate the quality of the lattice results one
should keep in mind that the LOPT results, see again 
Fig. \ref{fig:D} -- 
to be achieved in the extremely high
temperature limit -- would be
\begin{equation}
D^{LOPT} = \frac { 6 \pi}{g^4  T log (2 T / m_D)} \simeq 80
\end{equation} 
On the scale of the huge range of allowed
$D$ values, from the extreme strongly coupled to the
free plasma, the agreement between the two lattice calculations,
carried out with different strategies and actions, is 
impressive. 

\section{Summary}
A precision era for high temperature QCD has started,
with results whose accuracy  is often comparable with the
one achieved at zero temperature. 
Staggered  bulk thermodynamics in the continuum and at the physical point, 
 well into the current LHC region is in good control,
and  residual discrepancies seem minor.
Wilson fermions are becoming competitive, with results in
the continuum limit (albeit still with largish masses) for bulk
thermodynamics, and in the critical region. 
New analytic studies in the high temperature phase compare remarkably
well with  numerical results, down to unexpectedly low temperatures. 
Chiral fermions are coming of age, with new, continuously developing
high quality results for domain wall and overlap fermions.
Good chiral properties together with new theoretical insight trigger 
further activity on chiral and axial symmetries and their
spectral properties.
Details of QCD dynamics prove to be 
important and delicate close to the phase transition: the 
fate of the axial anomaly, the order of the transition
for $N_f = 2$,the  response to a magnetic field, and the role of a $\theta$ 
therm are still unclear. 
New  lattice results for transport and quarkonia 
have been presented which compare well with RHIC and LHC experiments.
The (strongly coupled) Quark Gluon Plasma is heavily investigated,
either in QCD and by use of deformed QCD and other model field theories, 
and confirms its role of inspiring   theoretical laboratory. 

\vspace{0.1cm}
\noindent
{\bf Acknowledgements}

I am most grateful to the many colleagues who have shared their results and notes. I have prepared this talk during a visit in Bielefeld in Summer 2012, were I have greatly enjoyed, and benefited from many discussions:   in particular I wish to thank Edwin Laermann, York Schroeder, Alexi Vuorinen and Christian Schmidt. Finally, it is a pleasure to thank the organisers for a very  interesting and smoothly run meeting. 

\end{document}